 \documentclass[a4paper,12pt]{article}
\usepackage[utf8]{inputenc}
\usepackage{setspace}

\title{Bayesian inference with tmbstan for a state-space model with VAR(1) state equation}

\author{Yihan Cao\thanks{Centre for Biodiversity Dynamics, Department of Mathematical Sciences, Norwegian University of Science and Technology, 7491 Trondheim, Norway}\and Jarle Tufto\footnotemark[1]}

 \usepackage{amsmath}
 \usepackage[margin=1in]{geometry}
 \usepackage[utf8]{inputenc}
 \usepackage[T1]{fontenc}
 \usepackage{textcomp}
 \usepackage{hyperref}
 \usepackage{graphicx}
 \usepackage{mwe}
 \usepackage{xr}
 \usepackage{booktabs}
 \usepackage{multirow}
\usepackage{amsmath,amsfonts,amsthm,bm} 
  \usepackage{caption}
 \usepackage[nodoi]{apacite}
 \usepackage{subfig}

 \newcommand{\Pois}{\operatorname{Poisson}}
\begin{document} 
  \date{}
  \maketitle
 
\section{Introduction}
Both frequentist and Bayesian statistical inference have been used for investigating ecological processes.
In the frequentist framework, Template model builder \shortcite<TMB,>{kristensen2016tmb}, an R package developed for fast fitting complex linear or nonlinear mixed models, has gained the popularity recently, especially in the field of ecology which usually involves in modeling complicated ecological processes \shortcite<for example>{cadigan2015state,albertsen2016choosing,auger2017spatiotemporal}.  The combination of reverse-mode automatic differentiation and Laplace approximation for high-dimension integrals makes parameter estimation with TMB very efficient even for non-Gaussian and complex hierarchical models. TMB provides a flexible framework in model formulation and can be implemented even for statistical models where the predictor is nonlinear in parameters and random effect. However, the lack of capability of working in the Bayesian framework has hindered the adoption of it for Bayesians.

Within the Bayesian framework, the software package \textit{Stan} \shortcite{gelman2015stan}, a probabilistic programming language for statistical inference written in C++ attracts peoples attention. It uses the No-U-Turn Sampler (NUTS) \cite{hoffman2014no}, an adaptive extension to Hamiltonian Monte Carlo \cite{neal2011mcmc}, which itself is a generalization of the familiar Metropolis algorithm, to conduct sampling more efficiently through the posterior distribution by performing multiple steps per iteration. Stan is a valuable tool for many ecologists utilizing Bayesian inference, particularly for problems where BUGS \shortcite{lunn2000winbugs} is prohibitively slow \shortcite{monnahan2017faster}. As such, it can extend the boundaries of feasible models for applied problems, leading to a better understanding of ecological processes. Fields that would likely benefit include estimation of individual and population growth rates, meta‐analyses and cross‐system comparisons, among many others.

Combining the merits of TMB and Stan, the new software package \textit{tmbstan} \cite{monnahan2018no} which provides MCMC sampling for TMB models was developed. This package provides ADMB and TMB users a possibility for making Bayesian statistical analysis when prior information on the unknown parameters is available. From the user’s perspective, it implements NUTS sampling from a target density proportional to the product of marginal likelihood (computed by TMB or Stan) and the prior density specified by the user. The user has the flexibility to decide which random effects are integrated out via the Laplace approximation in TMB and then the TMB model is passed to function Stan in the RStan package so that the rest of the parameters are integrated by Stan. This methodology might therefore potentially be more computationally efficient than using MCMC alone to integrate out all parameters. \citeA{monnahan2018no} introduced the \textit{tmbstan} package, applied it to simulation studies and compared its capabilities (computational efficiency and the accuracy of Laplace approximation) with other platforms such as ADMB and TMB. 

However, it is unclear that if Bayesian inference with arbitrary prior distribution implemented with Stan would perform comparatively with frequentist inference when modeling complex ecological processes. It is also unclear that when using tmbstan, if using the Laplace approximation to integrate latent variables is more computationally efficient than handling all latent variables via MCMC. In the case studies in \citeA{monnahan2018no}, Laplace approximation turned out to reduce the computational efficiency of MCMC. Another issue arose in the case studies is that the Laplace approximation to the integration of random effects is not accurate to a degree and this could lead to biased parameter estimates or uncertainties in parameter estimation. To gain more insights on these issues, in this paper we conduct simulation studies and a case study in the context of modeling fluctuating and auto-correlated selection with state-space models (SSM).  These forms of models are more generally increasingly used in ecology to model time-series such as animal movement paths and population dynamics \cite<for example>{cadigan2015state,albertsen2016choosing,auger2017spatiotemporal}. Furthermore, following \citeA{cao2019time}, we also use order-1 vector autoregressive model (VAR(1)) to model the unobserved states, which in our study are temporally fluctuating and potentially auto-correlated height, width and location of a Gaussian fitness function. This also allows us to make a further investigation into the issue of underestimation of the auto-correlation parameter in auto-regressive models shown in \shortciteA{Chevin2015} and \citeA{cao2019time}. 

Through the simulation and empirical studies, our paper aims to (1) compare estimates between frequentist inference and Bayesian inference under different simulation schemes; (2) investigate how the choice of prior influence Bayesian inference; (3) compare the computational efficiency of MCMC with and without integrating out some of the random effects via Laplace approximation.

\section{Methodology}
\subsection{Model formulation}
We consider a typical ecological process, the fluctuating selection in a bird species, the great tit (\textit{Parus major}). We conduct the study in the context of temporally changing selection on the laying date with the number of fledglings as the fitness component, but it can be generalized to any episode of viability or fertility selection, or to overall selection through lifetime fitness.  The discrete nonnegative variable, number of fledglings, is best modelled by distributions such as Poisson, or zero-inflated Poisson \cite<for example>{Chevin2015,cao2019time}. Within the framework of generalized linear models, the expected value 
of response variable is commonly linked to the linear predictors of biologically interest by logarithm. When both linear and quadratic effects of the traits are included, this leads to a Gaussian model of stabilizing selection. In this study, the number of fledglings in a specific brood is assumed to be Poisson distributed, $X_{i}  | w_{i} \sim \Pois(w_{i})$, where $i$ indicates the breeding event. The fitness (the expected number of fledglings $w_{i}$ ) of individuals with phenotype $z_i$ is then given by
\begin{equation} \label{eq:1}
\begin{aligned}
 \ln w_{i} &= \eta_{t}^{(\alpha)} - 
\frac{(z_{i}-\eta_{t}^{(\theta)} )^2}{2(e^{\eta_{t}^{(\omega)} })^2},\\
\end{aligned}
\end{equation}
where $\eta_{t}^{(\alpha)}$, $\eta_{t}^{(\theta)}$ and  $e^{\eta_{t}^{(\omega)}}$ ($e$ based to guarantee positive) are
parameters determining the logrithm of maximum fitness, optimum laying date and width of the fitness function 
in year $t$ respectively.
We further model $\eta_{t}^{(\alpha)}$, $\eta_{t}^{(\theta)}$ and $\eta_{t}^{(\omega)}$, 
the three stochastic processes as following:
\begin{equation} \label{eq:2}
        \begin{aligned}
\eta_{t}^{(\alpha)} &=\mu_{\alpha}+ \sigma_{\alpha}\alpha_t,\\
\eta_{t}^{(\theta)} &=\mu_{\theta}+ \sigma_{\theta}\theta_t,\\
\eta_{t}^{(\omega)} &=\mu_{\omega}+ \sigma_{\omega}\omega_t.
 \end{aligned}
        \end{equation}
The elements of vector $\mathbf{\mu}=(\mu_{\alpha}, \mu_{\theta}, \mu_{\omega})^T$ are the means
of the three processes.  The  stochastic processes
\(\alpha_{t}\), \(\theta_{t}\), \(\omega_{t}\) are
assumed to be multivariate normal distributed 
$(\alpha_t, \theta_t, \omega_t)^{T} \sim  \mathbf{N}_3(\mathbf{0}, \mathbf{\Gamma_0})$
with $\mathbf{\Gamma_0}=\begin{bmatrix}
1 &  \rho_{\alpha,\theta} &\rho_{\alpha,\omega}   \\ 
 \rho_{\alpha,\theta}& 1 & \rho_{\theta,\omega}  \\ 
 \rho_{\alpha,\omega}& \rho_{\theta,\omega} & 1
\end{bmatrix}$,
where $\rho_{\alpha,\theta}$, $\rho_{\alpha,\omega}$ and $\rho_{\theta,\omega}$ indicate the correlations and are assumed  to be mutually independent. $(\alpha_t, \theta_t, \omega_t)^{T}$ are further assumed to follow a first-order vector autoregressive (VAR(1)) process as below:

\begin{equation} \label{eq:3}
\left[\begin{matrix}
\alpha_t \\
\theta_t\\
\omega_t
\end{matrix}\right]
=
\mathbf{\Phi}
\left[\begin{matrix}
\alpha_{t-1} \\
\theta_{t-1} \\
\omega_{t-1}
\end{matrix}\right]
+
\mathbf{w}_t, 
\end{equation}
where \(\mathbf{\Phi}\) is $3\times3$ transition matrix and \(\mathbf{w}_t\) is a 3-dimentional vector of white noise. The 
covariance matrix of \(\mathbf{w}_t\) is calculated as $\mathbf{\Gamma_0}-\mathbf{\Phi}\mathbf{\Gamma_0}\mathbf{\Phi}$. Correlations
between the elements of \(\mathbf{w}_t\) are determined by both \(\mathbf{\rho}= (\rho_{\alpha,\theta}, \rho_{\alpha,\omega}, \rho_{\theta,\omega})\)  and $\mathbf{\Phi}$. If  $\mathbf{\rho}$ is  $\mathbf{0}$
vector and \(\mathbf{\Phi}\) is diagonal, then \(\mathbf{w}_t\) reduces to be three independent and identically distributed  white noise processes. In this case,
\(\alpha_t\), \(\theta_t\) and \(\omega_t\) simplify to three independent first-order autoregressive (AR(1)) processes. 
If $\rho$ is  $\mathbf{0}$ and all entries of \(\mathbf{\Phi}\) are zero, both $(\alpha_t, \theta_t, \omega_t)^{T}$ and  \(\mathbf{w}_t\) reduce to
three independent and identically distributed white noise processes.  In any case, our non-centered parameterization implies that the standard deviation
of $\alpha_t$, $\theta_t$ and  $\omega_t$ is only determined by $\sigma_{\alpha}$, $\sigma_{\theta}$ and $\sigma_{\omega}$ respectively.
We expect the non-centered parameterization yields simpler posterior geometries \cite{betancourt2015hamiltonian} and 
will be much more efficient in terms of effective sample size when there is not much data \cite[chapter~20]{stan2018stan}.

It is worth mentioning that one objective of this study is to provide another case study beyond the ones in \citeA{monnahan2018no}.
Therefore, even though $\alpha_t$, $\theta_t$ and $\omega_t$ are assumed to be VAR(1) in the model,
in the simulation study we consider only AR(1) $\theta_t$ and white noise of $\alpha_t$ and $\omega_t$.
The alternative simulation studies in which $\alpha_t$, $\theta_t$ and $\omega_t$ are formulated as
other possible stochastic processes can be conducted similarly and exhaustively, but that is an enormous amount of work
in one single study. When $\alpha_t$, $\theta_t$ and $\omega_t$ are assumed to be VAR(1), one caution to be taken is that all the eigenvalues of $\mathbf{\Phi}$ must lie in the unit circle to guarantee the VAR (1) process to be stationary \cite{wei2006time}.
At last, in the simulation study, we assume that the model structure is known, which means that we already know
$\theta_t$ is AR(1) process since the aim of the study is not to explore the structure of the true model.

\subsection{Prior distribution}
The priors are assumed to be independent to each other
$\pi(\mathbf{\mu,\Phi,\Sigma})=\pi(\mathbf{\mu})\pi(\mathbf{\Phi})\pi(\mathbf{\Sigma})$.
We take a normal $N(\mathbf{m},q\mathbf{I}_3)$ prior distribution for the process mean vector $\mathbf{\mu}=(\mu_\alpha,\mu_\theta,\mu_\omega)$
and input weak prior information on the process mean by taking $\mathbf{m} = \mathbf{0}$ and $q = 100$.
Since in this study we assume constant $\eta_t^{(\alpha)}$ and $\eta_t^{(\omega)}$,
$\phi_{\theta,\theta}$ is the only non-zero entry in $\mathbf{\Phi}$. We used truncated normal prior on $\phi_{\theta,\theta}$ since it outperforms Jeffreys’ prior \cite{jeffreys1961theory}, g prior \cite{zellner1986assessing} and natural conjugate prior
\cite{raiffa1961applied} in terms of posterior sensitivity using Highest Posterior Density Region (HPDR) criterion concluded from the simulation study in \shortciteA{karakani2016bayesian}.
\shortciteA{lei2011bayesian} also uses truncated normal distribution as subjective prior for the auto-regressive parameter in its AR (1) model.
The mean and standard deviation of the truncated normal distribution are arbitrarily set to be 0 and 0.5 respectively.

For the variance of the error term $\sigma_\theta^2$ ($\sigma_\alpha^2$ and $\sigma_\omega^2$ are assumed to be zero), two priors are used:\\
(1) half-Cauchy (0, 10) prior on $\sigma_\theta$ (Prior1); \\
(2) lognormal (1, 0.5) prior on  $\sigma_\theta$ (Prior2). \\
These two priors are referred to Prior1 and Prior2 respectively in the rest of this paper.
It is worth mentioning that we also tested uniform prior on $log(\sigma_\theta)$ (non-informative improper prior which
equals to $1/\sigma$ prior on $\sigma$ \cite{gelman2006prior}) and inverse-gamma (1, 1) prior on $\sigma_\theta^2$ 
(non-informative proper prior, also illustrated in \cite{gelman2006prior}), but both of them render an issue that the sampler traps 
in a subspace of the whole parameter space of $log(\sigma_\theta)$ and results in numerous divergent transitions. 
It was potentially caused by the posterior becoming improper and consisting of a mode and an infinite low-posterior-density ridge extending to infinity as illustrated in \citeA{tufto2012estimating}. We thus in this study only consider the two proper informative 
priors (Prior1 and Prior2), while more information on the MCMC with inverse-gamma (1, 1) prior on $\sigma_\theta^2$ 
is given in Supporting Information.

Note also that the scale parameters  $\log(\sigma_\theta)$ is declared in the TMB template in the logarithmic format, but the half-Cauchy prior and lognormal prior contributed to the total likelihood with the log density in terms of $\sigma_\theta$ and for inverse-gamma prior, it is in terms of $\sigma_\theta^2$, where $\sigma_\theta$ is a positive transform $\sigma= e^{log\sigma}$. Therefore, Jacobian adjustment (see chapter 20.3  in \citeA{stan2018stan} for Jacobian adjustment) was conducted by adding $log\sigma_\theta$ to the total likelihood when half-Cauchy prior and lognormal prior are used. When testing inverse-gamma prior, it was $\log 2+2\log\sigma_\theta$ added to the total likelihood. 

\subsection{Software implementation}
The model is formulated with C++ and passed to TMB for frequentist inference.
The model objective (fn) and gradient (gr) functions are fed to optimization 
function \textit{nlminb} with default setting to optimize the objective function.

For Bayesian inference, the TMB model objective and gradient functions 
are passed to \textit{tmbstan} which uses the \textit{stan} function and 
executes the No-U-Turn sampler (NUTS) algorithm by default to sample. 
Currently the other options are "HMC" (Hamiltonian Monte Carlo), 
and "Fixed\underline{\space}param". We ran the simulation study on a multicore computing server with enough RAM to avoid swapping to disk.
The number of warmup iterations to be excluded when computing the summaries
is set to 1000 and for total sample length, it is 3000. We thin each chain to every second sample and set the value adapt\underline{\space}delta to 0.95, which is the average proposal acceptance probability Stan aims for during the adaption (warmup) period.
We set a seed for each simulation including data set and tmbstan to make sure all the simulation results are reproducible.

Divergent transitions during sampling may occur due to a large step size in the sampler or a poorly parameterized model, meaning that the iteration of the MCMC sampler runs into numerical instabilities \cite{carpenter2017stan} and thus inferences will be biased. RStan team suggested that the problem may be alleviated by increasing the adapt\underline{\space}delta parameter
(gives a smaller step size), especially when the number of divergent transitions is small \cite{team2018rstan}. In our simulation studies, we
find it difficult to completely avoid divergent transitions across all data sets even though adapt\underline{\space}delta is increased to 0.95. Similar to \citeA{fuglstad2019intuitive}, we thus removed simulations where 0.1\% or more divergent transitions in the iterations after warmup occur during the inference to avoid reporting biased results. 

It is worth mentioning that the execution of Markov chains can be done in parallel.  While the default of RStan is to use 1 core, the RStan team recommended to set it to as many processors as the hardware and RAM allow and at most one core per chain \cite{team2018rstan}.  The simulations we run are done with a server that has 28 available cores. We thus set the number of cores to be 4 for the 4 Markov chains. However, since R is single threaded and for frequentist inference, optimization algorithm used in R function "nlminb" only uses one core of CPU, we thus only compare the computational efficiency between 
tmbstan with and without Laplace approximation and ignore the computational efficiency with "nlminb" to ensure fair comparisons.

\section{Simulation scheme and results}
\subsection{Simulation scheme}
All the data  simulated are in natural units and considered to be biologically realistic according to the empirical studies of 
natural birds populations \shortcite<e.g.>{grant2002unpredictable, vedder2013quantitative}.
Samples were modeled from a population undergoing stabilizing selection with AR(1) $\theta_t$, fixed $\eta_{t}^{(\alpha)}$ and
$\eta_{t}^{(\omega)}$. Vector $\mathbf{\mu}=(\mu_{\alpha}, \mu_{\theta}, \mu_{\omega})^T$ is set to $(2,20,3.5)$.
The autocorrelation $\phi_{\theta,\theta}$ is set to 0.1, 0.4 and 0.7 (only positive values considered since the estimate of auto-correlation 
in temporal optimal laying date is positive, for example 0.3029 in \citeA{Chevin2015} and 0.524 in \citeA{cao2019time}), the variance of 
fluctuating optimal laying date $\sigma_\theta$ is set to 20. 

For each value of $\phi_{\theta,\theta}$, $tmax =25$ or 50 time points were simulated and for each time point
the sample size was drawn from a Poisson distribution with mean $n = 25$, 50 or 100 individuals. We considered 
four combinations of $tmax$ and $n$, which are $(tmax=25, n=50)$, $(tmax=25, n=100)$, $(tmax=50, n=25)$ and $(tmax=50, n=100)$. These four combinations are refered as simulation setting 1, 2, 3, 4 respectively in the following sections.
Similar to \citeA{cao2019time}, we neglected response to selection and used the same normal distribution  for simulating individual phenotype each year. The phenotypic standard deviation before selection $\sigma_z$ was set to 20, such that the strength of stabilizing selection $S = \sigma^2_z/(e^{\eta_{t}^{(\omega)} })^2 + \sigma^2_z $ \cite<e.g.>{Chevin2015}  was 0.267. For each individual, its fitness was computed from its phenotype using equation \eqref{eq:1}, and its actual number of offspring 
was then drawn from a Poisson distribution with mean $w_t(z)$.

\subsection{Frequentist vs. Bayesian estimates}
 The results of one single simulation obtained from maximum likelihood in the frequentist framework are compared with those from \textit{tmbstan}.  
 The summaries of the estimates with tmbstan are computed after dropping the warmup iterations and merging the draws from all the four chains.
 The frequentist and Bayesian estimates with different sample sizes and $\phi_{\theta,\theta}=0.4$
 are shown in Table \ref{table:1}, the estimates with other values of auto-correlation in $\theta_t$ ($\phi_{\theta,\theta}$ =0.1 and 0.7) can be found in Supporting Information.
 
 \begin{table}[]
\caption{Frequentsit and Bayesian estimates (standard errors) from the model with AR(1) $\theta_t$, autocorrelation in $\theta_t$ $\phi_{\theta,\theta}=0.4$, and different sample sizes ($(tmax=25, n=50)$, $(tmax=25, n=100)$, $(tmax=50, n=25)$ and $(tmax=50, n=100)$) from one realization of the simulation. For each sample size setting, the number of divergent transitions in the MCMC is also reported and is used as 
a measure of stability of the inference scheme. MLE stands for maximum likelihood estimate, Prior1 and Prior2 represent half-Cauchy (0, 10) and lognormal (1, 0.5) prior respectively. }
\label{table:1}

\begin{tabular}{|l|l|l|l|l|} \hline 
\multicolumn{5}{|c|}{$\phi_{\theta,\theta}=0.4$, $tmax=25$, $n=50$} \\ \hline 
Parameters & True value & MLE & Prior1 & Prior2  \\ \hline 
no. divergent transitions & NA & NA &1  &  0 \\ \hline 
$\mu_\alpha$ & 2 &2.017(0.015) & 2.017(0.015) & 2.016(0.015) \\ \hline 
$\mu_\theta$ & 20 &18.5(3.7) & 18.3(5.1) & 18.5(3.7) \\ \hline 
$\mu_\omega$ & 3.5 &3.472(0.028) & 3.475(0.028) & 3.469(0.028) \\ \hline 
$\phi_{\theta,\theta}$ & 0.4 &0.14(0.20) & 0.23(0.23) & 0.16(0.18) \\ \hline 
$log\sigma_\theta$ & 2.996 &2.77(0.15) & 2.88(0.19) & 2.70(0.14) \\ \hline 
\multicolumn{5}{|c|}{$\phi_{\theta,\theta}=0.4$, $tmax=25$, $n=100$} \\ \hline 
Parameters & True value & MLE & Prior1 & Prior2  \\ \hline 
no. divergent transitions & NA & NA &2  &  0 \\ \hline 
$\mu_\alpha$ & 2 &1.995(0.011) & 1.995(0.012) & 1.995(0.012) \\ \hline 
$\mu_\theta$ & 20 &20.2(8.7) & 18.3(17.5) & 20.1(7.4) \\ \hline 
$\mu_\omega$ & 3.5 &3.506(0.022) & 3.508(0.022) & 3.504(0.021) \\ \hline 
$\phi_{\theta,\theta}$ & 0.4 &0.50(0.17) & 0.59(0.18) & 0.46(0.13) \\ \hline 
$log\sigma_\theta$ & 2.996 &3.25(0.18) & 3.43(0.28) & 3.13(0.14) \\ \hline 
\multicolumn{5}{|c|}{$\phi_{\theta,\theta}=0.4$, $tmax=50$, $n=25$} \\ \hline 
Parameters & True value & MLE & Prior1 & Prior2  \\ \hline 
no. divergent transitions & NA & NA &0  &  0 \\ \hline 
$\mu_\alpha$ & 2 &1.974(0.015) & 1.974(0.015) & 1.973(0.015) \\ \hline 
$\mu_\theta$ & 20 &20.0(3.8) & 19.8(4.9) & 20.1(4.2) \\ \hline 
$\mu_\omega$ & 3.5 &3.520(0.032) & 3.523(0.032) & 3.515(0.031) \\ \hline 
$\phi_{\theta,\theta}$ & 0.4 &0.42(0.14) & 0.48(0.15) & 0.42(0.13) \\ \hline 
$log\sigma_\theta$ & 2.996 &2.84(0.13) & 2.92(0.16) & 2.79(0.13) \\ \hline 
\multicolumn{5}{|c|}{$\phi_{\theta,\theta}=0.4$, $tmax=50$, $n=100$} \\ \hline 
Parameters & True value & MLE & Prior1 & Prior2  \\ \hline 
no. divergent transitions & NA & NA &0  &  0 \\ \hline 
$\mu_\alpha$ & 2 &1.9865(0.0076) & 1.9864(0.0076) & 1.9861(0.0076) \\ \hline 
$\mu_\theta$ & 20 &20.7(3.9) & 20.0(5.0) & 20.7(4.1) \\ \hline 
$\mu_\omega$ & 3.5 &3.512(0.015) & 3.513(0.015) & 3.510(0.015) \\ \hline 
$\phi_{\theta,\theta}$ & 0.4 &0.41(0.13) & 0.47(0.15) & 0.41(0.12) \\ \hline 
$log\sigma_\theta$ & 2.996 &2.89(0.12) & 2.97(0.17) & 2.85(0.11) \\ \hline 
\end{tabular}

\end{table}

From Table \ref{table:1} we find that both frequentist and Bayesian inferences show good estimates for  $\mu_\alpha$ and $\mu_\omega$. It is interesting to see that the auto-correlation for $\theta_t$ is not always under-estimated under all settings (for example $(tmax=25, n=50)$), this can be also seen from the tables for parameter estimates in Supporting Information. Bayesian inference with Prior1 (half-Cauchy prior) generally reports smaller estimates of $\mu_\theta$ than MLE and Prior2 (lognormal prior) but larger estimates of $\phi_{\theta,\theta}$ and $log\sigma_\theta$. The estimates with MLE and Prior2 are close to each other while the estimates with Prior2 show fewer uncertainties for $\phi_{\theta,\theta}$ and $log\sigma_\theta$ implied by the smaller standard errors in the brackets. Prior2 also reports smaller estimates for $log\sigma_\theta$ compared with MLE and Prior1 since it puts very large weight on small values of the variance, as will be graphically demonstrated in section 3.4.  We also find that $\phi_{\theta,\theta}$ and $\log\sigma_\theta$ are difficult parameters to estimate since none of these three techniques can estimate them accurately across all the cases. However, the estimates are based on one realization of simulation, the discrepancy between estimates to the true value would vary from simulation to simulation. 

We also compare the estimates across the different sample sizes. We typically compare the estimates 
between setting $(tmax=25,n=50)$ and $(tmax=25, n=100)$, $(tmax=50, n=25)$ and $(tmax=50, n=100)$, $(tmax=25, n=100)$ and $(tmax=50, n=100)$.
We find that increasing the mean sample size at each time point does not necessarily increase the certainty of the estimates,
but the data set with increased time points $(tmax=50, n=100)$ contains more information on the parameters of interest and thus reports more certain
estimates compared with the data set with $(tmax=25, n=100)$. The same conclusion can be also drawn by making similar comparisons among the estimates in Table S1 and S2 in Supporting Information.

We can also find from Table \ref{table:1}, Table S1 and S2  from Supporting Information that the Bayesian inference with Prior1
in some cases report 1 or 2 divergent transitions while with Prior2 there are no divergent transitions reported. This implies that
the geometric shape of posterior likelihood with Prior1 is more challenging for sampling probably due to light tails and thus potentially leads to an incomplete exploration of the target distribution.

\subsection{Bias Plot}
\begin{figure}[h!]
\centering
\includegraphics[scale=0.32]{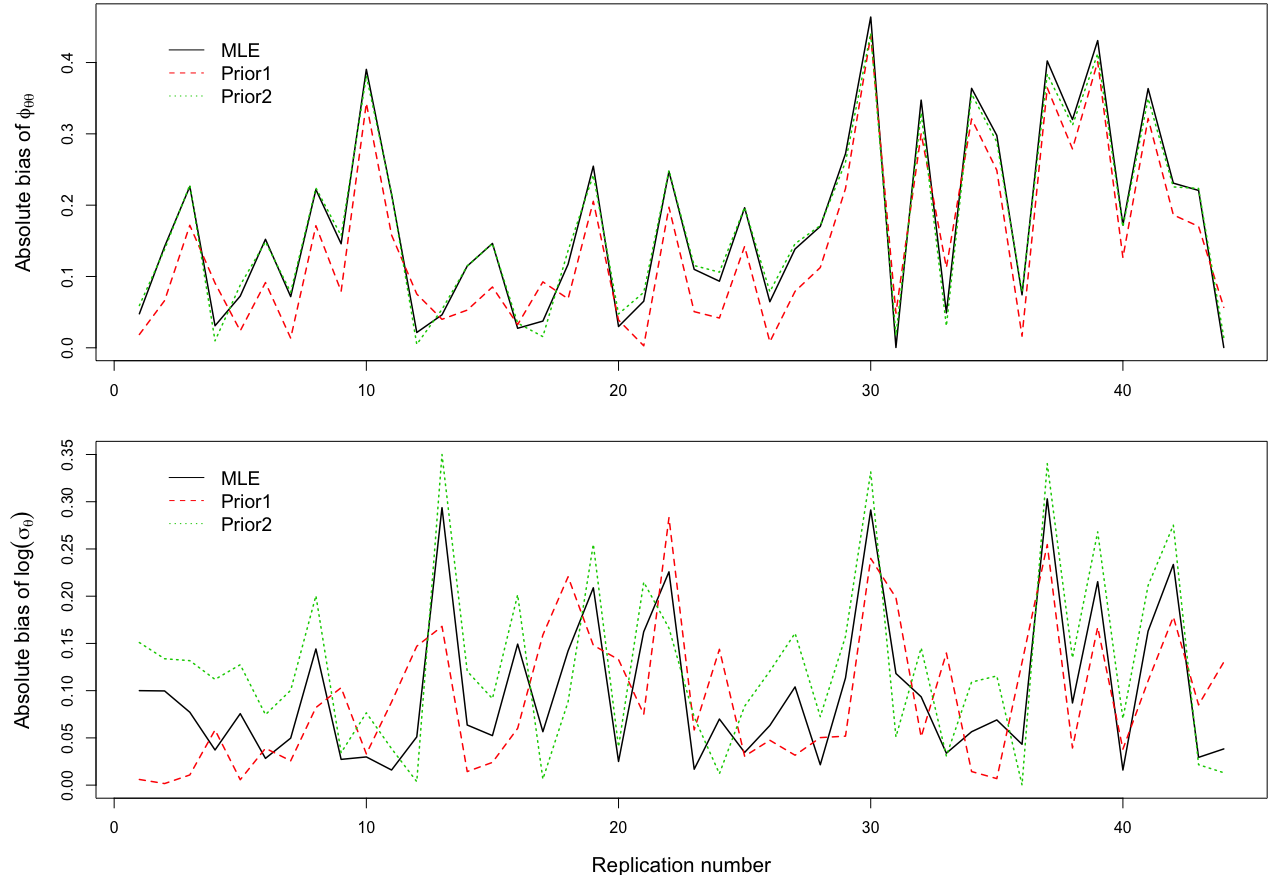}
\caption{Bias plots for the auto-regressive parameter $\phi_{\theta,\theta}$ (the upper plot) and for the scale parameter log$\sigma_{\theta}$ 
(the lower plot) respectively under the setting with time series length $tmax=50$, average annual sample size $n=25$, autocorrelation in $\theta_t$ $\phi_{\theta,\theta}=0.4$ and 44 replications (50 replications were conducted, among which 6 replications report 3 or more divergent transitions
for the MCMC of Bayesian inference and thus are removed from the analysis).}
\label{fig:bias}
\end{figure}

The comparison between the estimates in the last section is based on one realization of the simulation. To make comparisons of estimates over more realizations, the simulation was repeated 50 times under the setting of $(tmax=50, n=25)$. Due to divergent transitions, only 44 out of 50 replicates were kept and the replications with more than 0.1\% divergent transitions (in 2000 iterations) were excluded from the analysis. 
For the estimate of $\phi_{\theta,\theta}$ and $log\sigma_\theta$ in each replication, the bias was calculated in a frequentist framework as the absolute difference between the true value and the mean estimate from each inference technique. The absolute bias for $\phi_{\theta,\theta}$ and $log\sigma_\theta$ are graphically displayed in the upper and lower
plot in Fig.~\ref{fig:bias} respectively.  From the upper plot we find that in most replications, Bayesian inference with Prior1 slightly outperforms the frequentist inference and Bayesian inference with Prior2, the latter two reported very close estimates for $\phi_{\theta,\theta}$. One striking
thing is that the bias is close to or even larger than 0.4 for some replications, this suggests that the inferences report even negative estimates of $\phi_{\theta,\theta}$ and it again turns out to be a difficult parameter. In the lower plot, we can see no single inference technique stands out in estimating the scale parameter $log\sigma_\theta$.

\subsection{Prior-posterior distribution}
\begin{figure}%
    \subfloat[$(tmax=25,n=50)$.]{{\includegraphics[width=8.5cm]{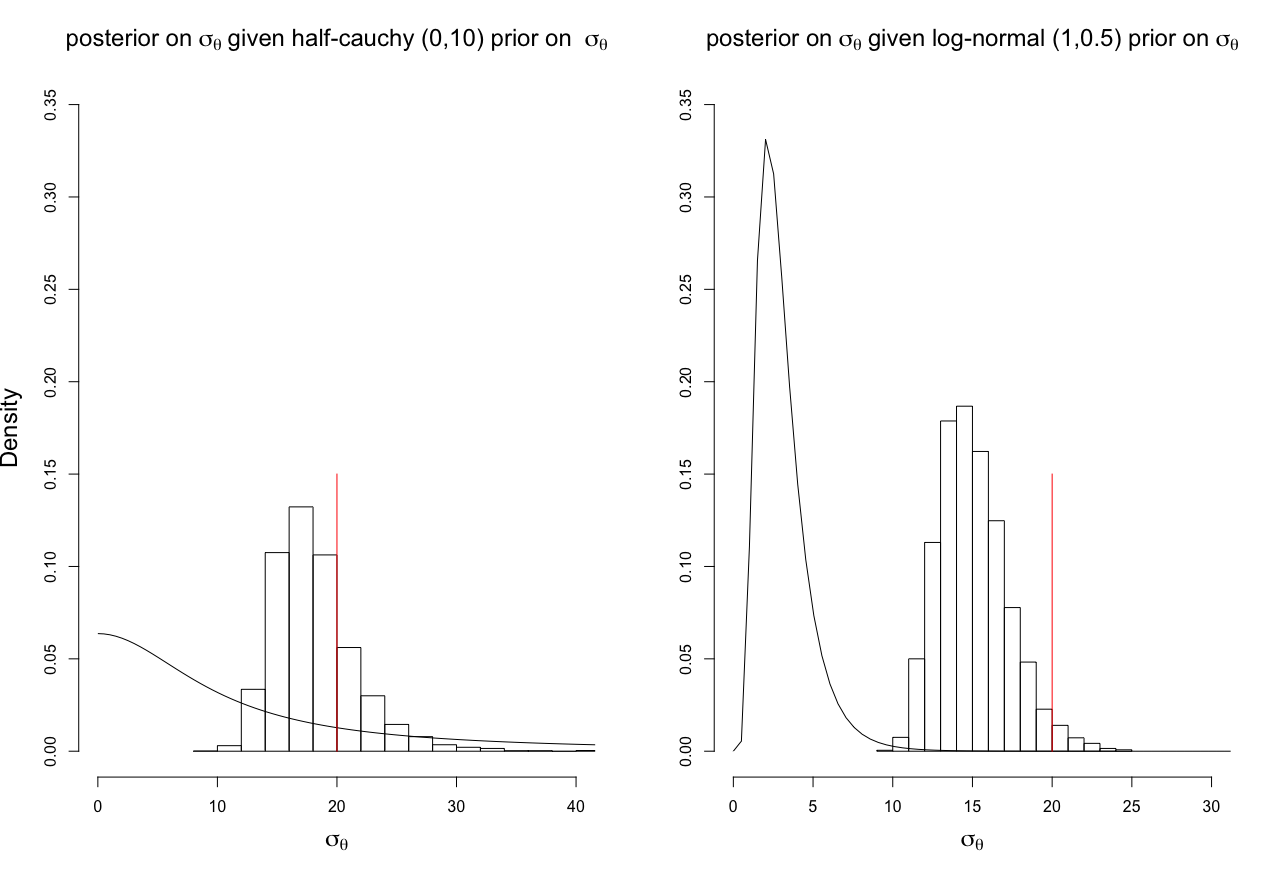}}}%
    \subfloat[$(tmax=25,n=100)$.]{{\includegraphics[width=8.5cm]{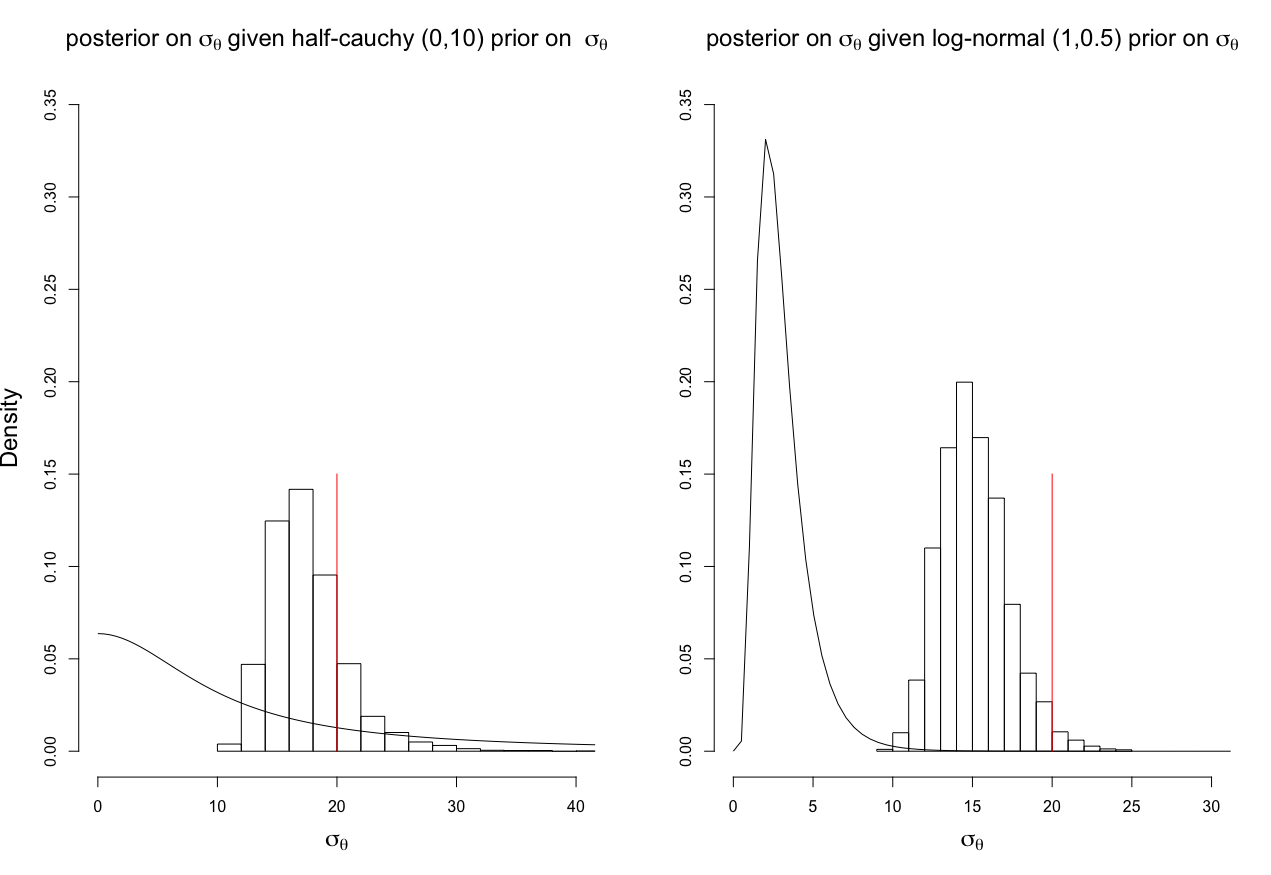}}}%
        \qquad
    \subfloat[$(tmax=50,n=25)$.]{{\includegraphics[width=8.5cm]{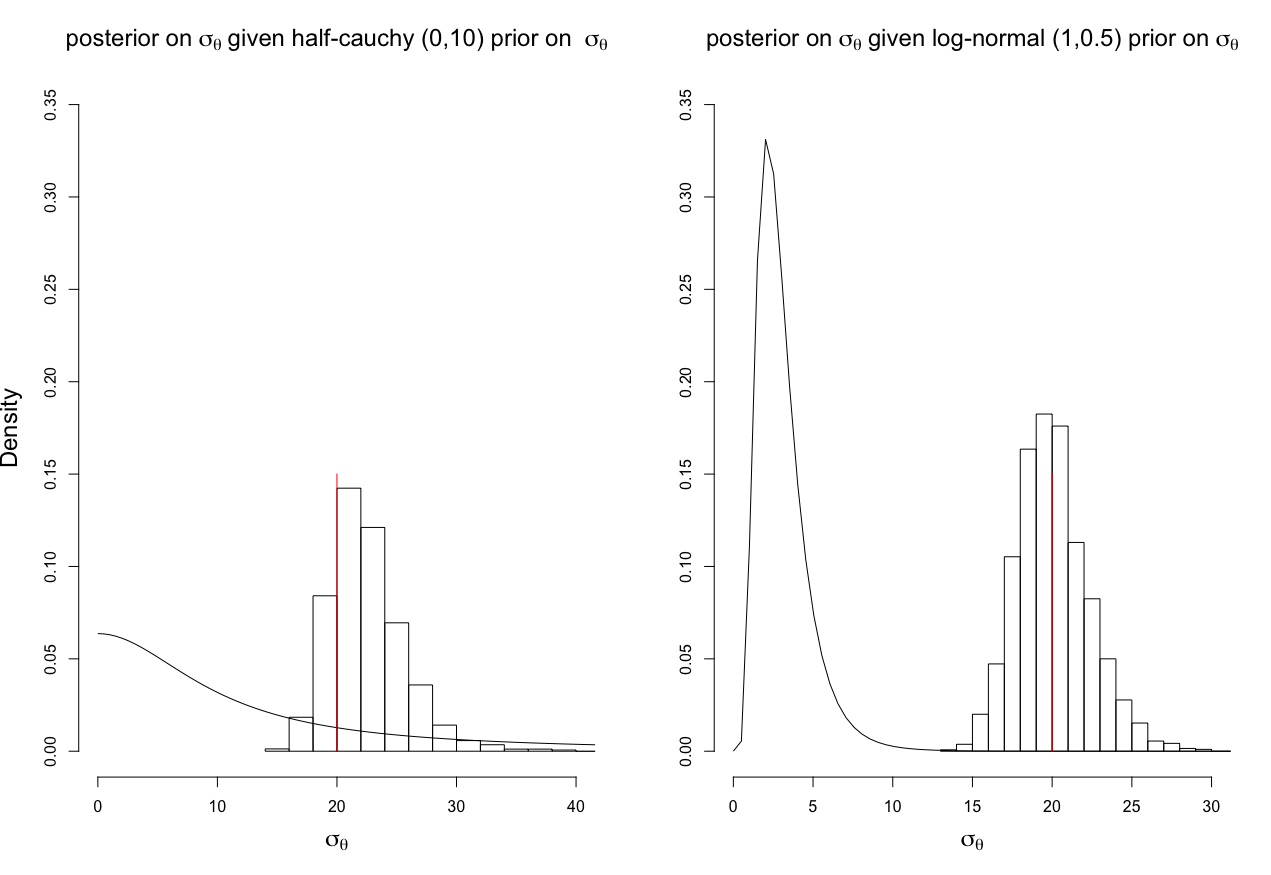}}}%
    \subfloat[$(tmax=50,n=100)$.]{{\includegraphics[width=8.5cm]{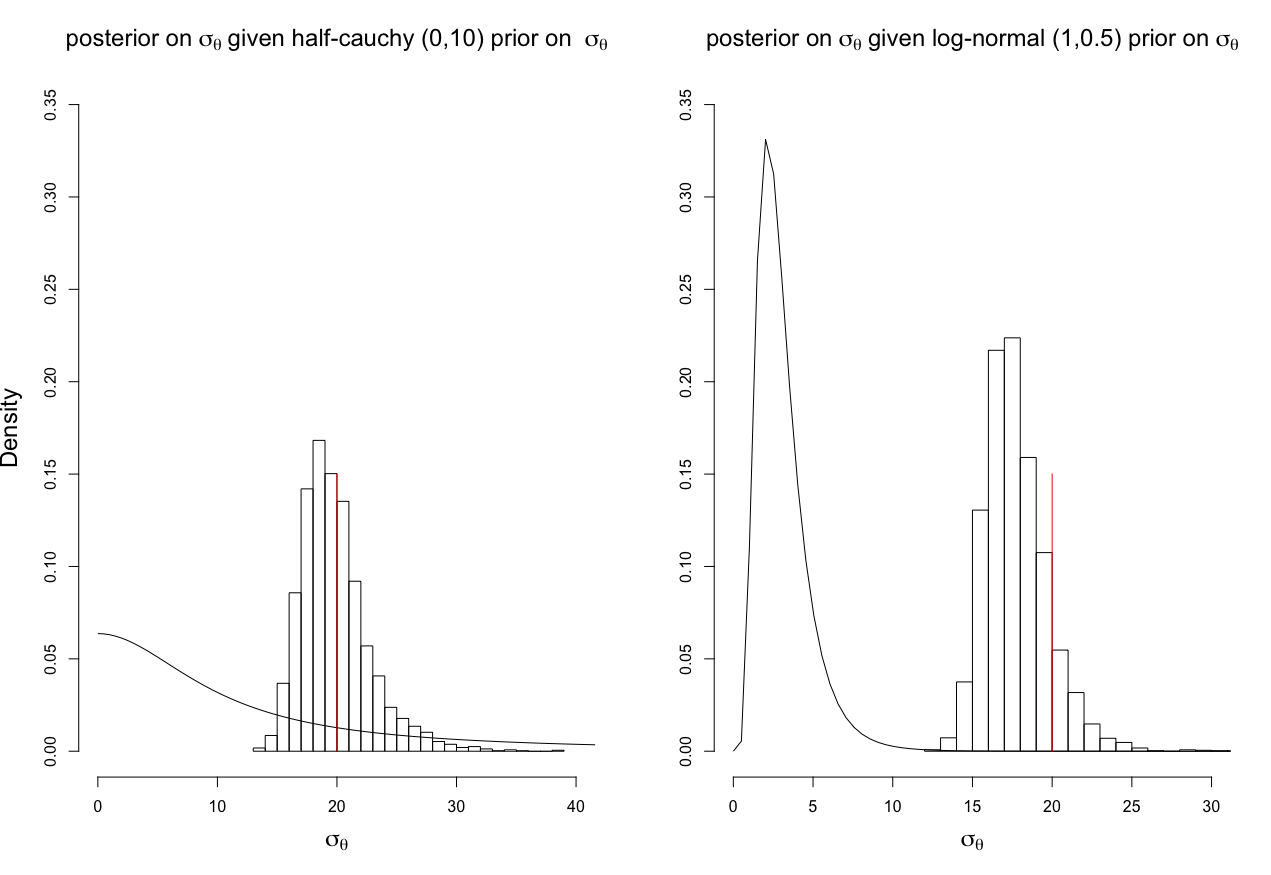}}}%
  \caption{Histograms of posterior samples of the scale parameter $\sigma_{\theta}$ 
   from models with two different prior distributions. Plot (a), (b), (c) and (d) correspond to sample size 
   setting $(tmax=25, n=50)$, $(tmax=25, n=100)$, $(tmax=50, n=25)$ and $(tmax=50, n=100)$ respectively.
   On each subplot, the left one 
   shows the histogram of posterior samples given half-Cauchy (0, 10) prior on $\sigma_{\theta}$
   and similarly, the right one displays the histogram of posterior samples given log-normal (1, 0.5)
   prior on $\sigma_{\theta}$. Overlain on each subplot (the solid black lines) is the corresponding prior density function.
   The red lines indicate the true value of $\sigma_{\theta}$.  Only  $\phi_{\theta,\theta}=0.4$
   was considered in the simulations.}
    \label{fig:pripost}%
\end{figure}

Fig.~\ref{fig:pripost} shows histograms of posterior samples of the scale parameter $\sigma_{\theta}$ 
from models with the two different prior distributions: half-Cauchy (0, 10) and log-normal (1, 0.5),
which are represented by solid lines in the left and right plot on each subplot respectively. The true value of $\sigma_{\theta}$ is indicated by a solid red line. Plot (a), (b), (c) and (d) correspond to 
setting $(tmax=25, n=50)$, $(tmax=25, n=100)$, $(tmax=50, n=25)$ and $(tmax=50, n=100)$ respectively. We can see from plot (a) that the priors are quite informative and pull the posteriors towards small
values away from the true value and this prior-domination is more clear with log-normal prior where
the prior distribution sharply peaks at 2. The domination is not mitigated even though the mean annual
sample size is increased to 100 as shown in plot (b). With the same total sample size in plot (c)
$(tmax=50, n=25)$ as that in plot (a) $(tmax=25, n=50)$, the posterior likelihoods in plot (c) are, however, not 
dominated by the priors. The prior-domination is also mitigated in plot (d) compared with plot (b) by increasing
the time points from 25 to 50.

Altogether, the informative log-normal prior pulls more of the posterior towards a narrower range of smaller parameter values especially when the number of time points in the data is small. The posterior samples are less dominated by the half-Cauchy prior in this case. Increasing the annual mean sample size does not necessarily lead to better identification of the small region of parameter space. Only the amount of time points is the matter for the likelihood to overwhelm the prior distribution and to dominate the posterior distribution. 

\subsection{Computational efficiency with and without Laplace approximation}
In \textit{tmbstan}, sampling can be performed with or without Laplace approximation for the random effects. 
It is possible to mix the Laplace approximation with MCMC by specifying \textit{laplace=TRUE}, such that the random effects are integrated with the Laplace approximation in TMB and other parameters (such as fixed effects and hyperparameters specifying the distribution of the random effects) are handled by the NUTS in Stan.
In the case studies in \citeA{monnahan2018no}, the Bayesian inference algorithms with Laplace approximation is less computationally efficient
than without Laplace approximation, where the efficiency is defined as the minimum effective sample size per second.  Following that definition, we calculated the efficiency of tmbstan with and without Laplace approximation with simulated data. Different from \citeA{monnahan2018no}, we did not consider the computational efficiency of Frequentist inference with the Laplace approximation, as explained in the last section. 
 
\begin{figure}%
    \centering
    \subfloat[Computational efficiency with Prior1.]{{\includegraphics[width=13cm]{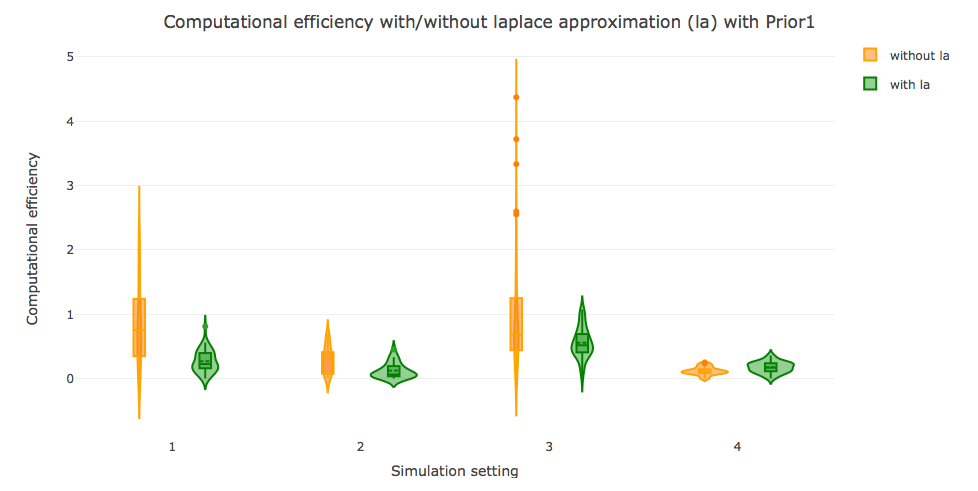}}}%
    \qquad
    \subfloat[Computational efficiency with Prior2.]{{\includegraphics[width=13cm]{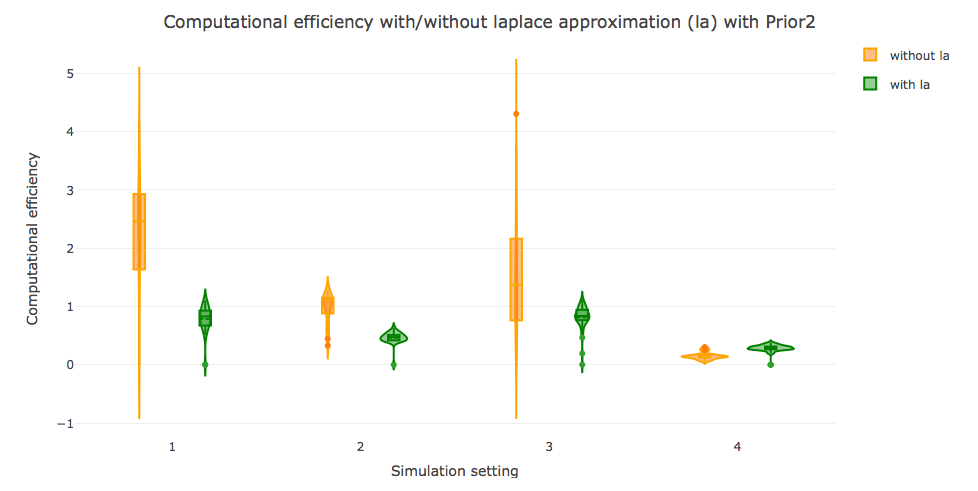}}}%
    \caption{Violin plots of computational efficiency (minimum effective sample size per second) without (orange) and with (green) Laplace approximation (la). The four settings
    on x axis correspond to sample size setting $(tmax=25, n=50)$, $(tmax=25, n=100)$, $(tmax=50, n=25)$ and $(tmax=50, n=100)$ respectively. 
    Plot (a) shows the computational efficiency of Bayesian inference with Prior1 and plot (b) with Prior2. Only $\phi_{\theta,\theta}=0.4$
    was used in simulations. Inside the violin plots are box plots showing the quantiles of 50 realized computational efficiencies. For each realization among the 50 simulations and across the settings, the same specifications in tmbstan are used.}
    \label{fig:efficiency}%
\end{figure}

In Fig.~\ref{fig:efficiency}, plot (a) displays violin plots of computational efficiency without (orange) and with (green) Laplace 
approximation (la) of Bayesian inference with Prior1 under different sample size settings. The setting 1, 2, 3, 4 on x axis stand for setting $(tmax=25, n=50)$, $(tmax=25, n=100)$, $(tmax=50, n=25)$ and $(tmax=50, n=100)$ respectively. Only $\phi_{\theta,\theta}=0.4$ was
considered and the divergent transitions were not taken into account. Inside the violin plots are box plots showing the quantiles 
of 50 realized computational efficiencies. Similarly, the  violin plots of computational efficiency with Prior2 are shown on plot (b).
We find from both plot (a) and (b) that Bayesian inference without Laplace approximation generally is more efficient under setting 1, 2, and
3, the outperformance is more manifest when the sample size is small $(tmax=25, n=50)$. However, when the sample size is increased to $(tmax=50, n=100)$, inference with Laplace approximation turns out to be slightly more efficient than that without Laplace approximation, the boxplots and violin plots also tend to be more compact under this setting. 

 Even though the technique in which the random effects are integrated out by Laplace approximation in TMB turns out to be less efficient in most settings, we still provide a counterexample from \citeA{monnahan2018no} in which the enabling of Laplace approximation is always less computationally efficient in the case studies.

\subsection{Laplace approximation check}
\begin{figure}%
    \centering
    \subfloat[$(tmax=25,n=50)$ with Prior2.]{{\includegraphics[width=8.5cm]{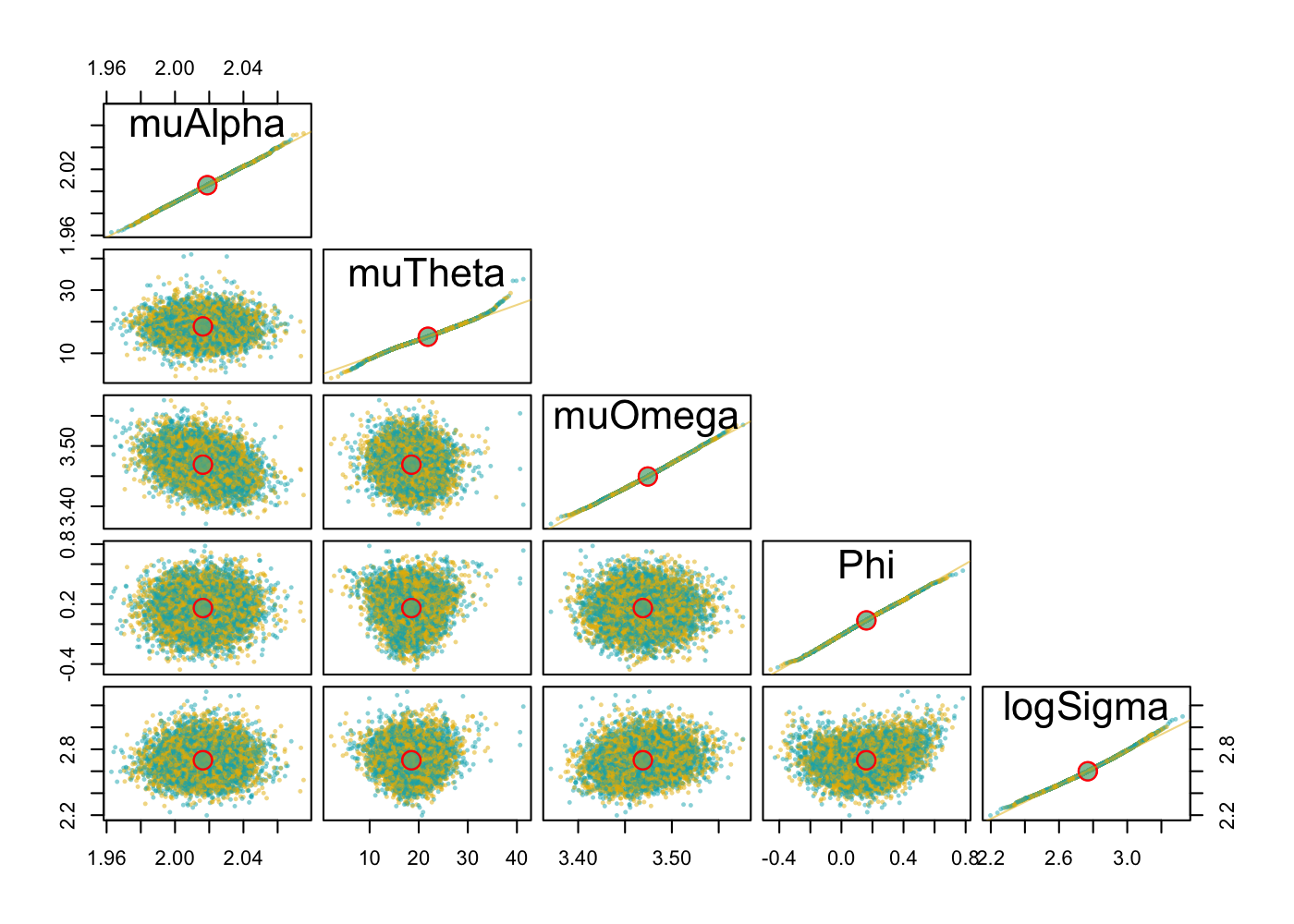}}}%
    \subfloat[$(tmax=25,n=100)$ with Prior2.]{{\includegraphics[width=8.5cm]{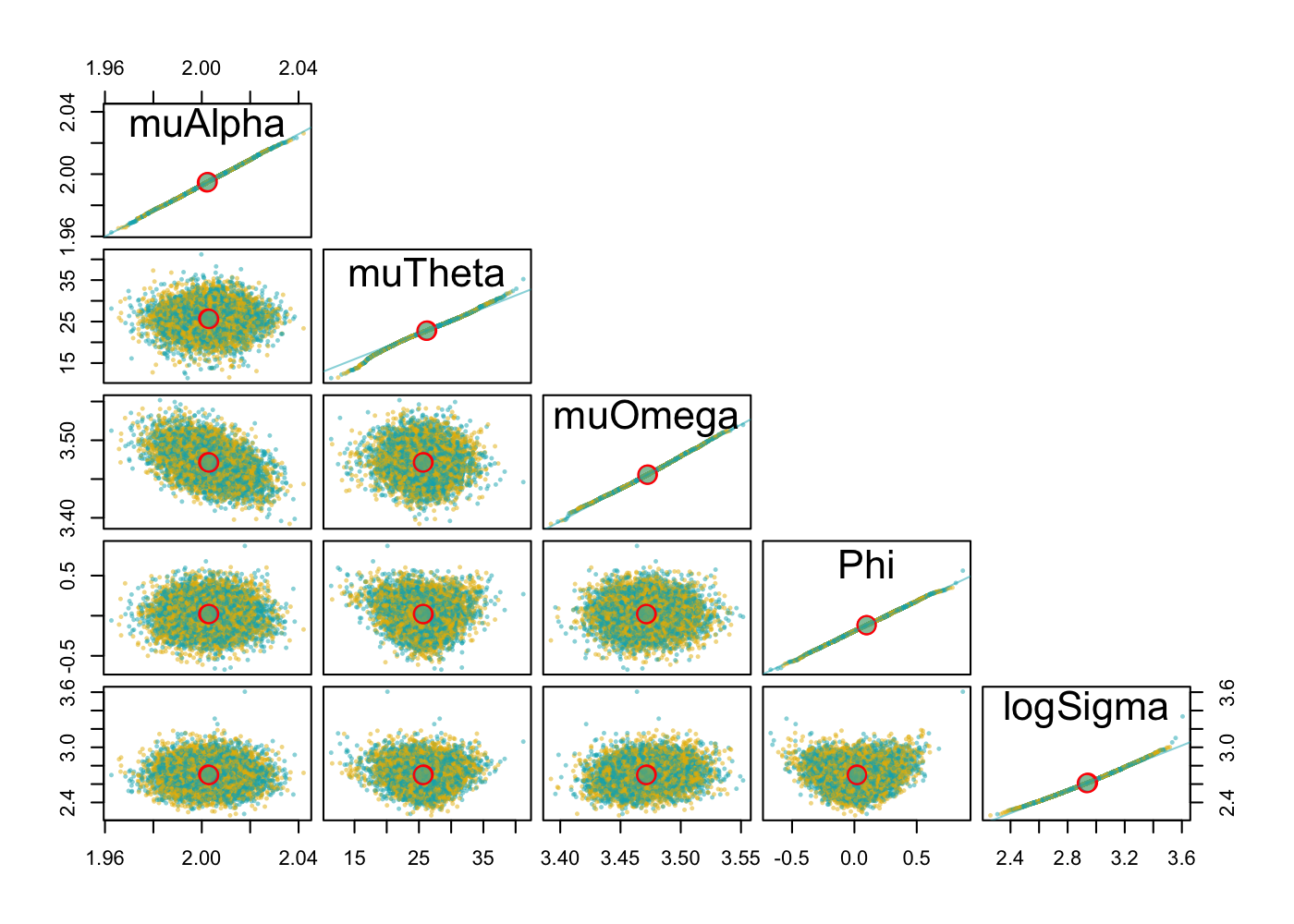}}}%
        \qquad
    \subfloat[$(tmax=50,n=25)$ with Prior2.]{{\includegraphics[width=8.5cm]{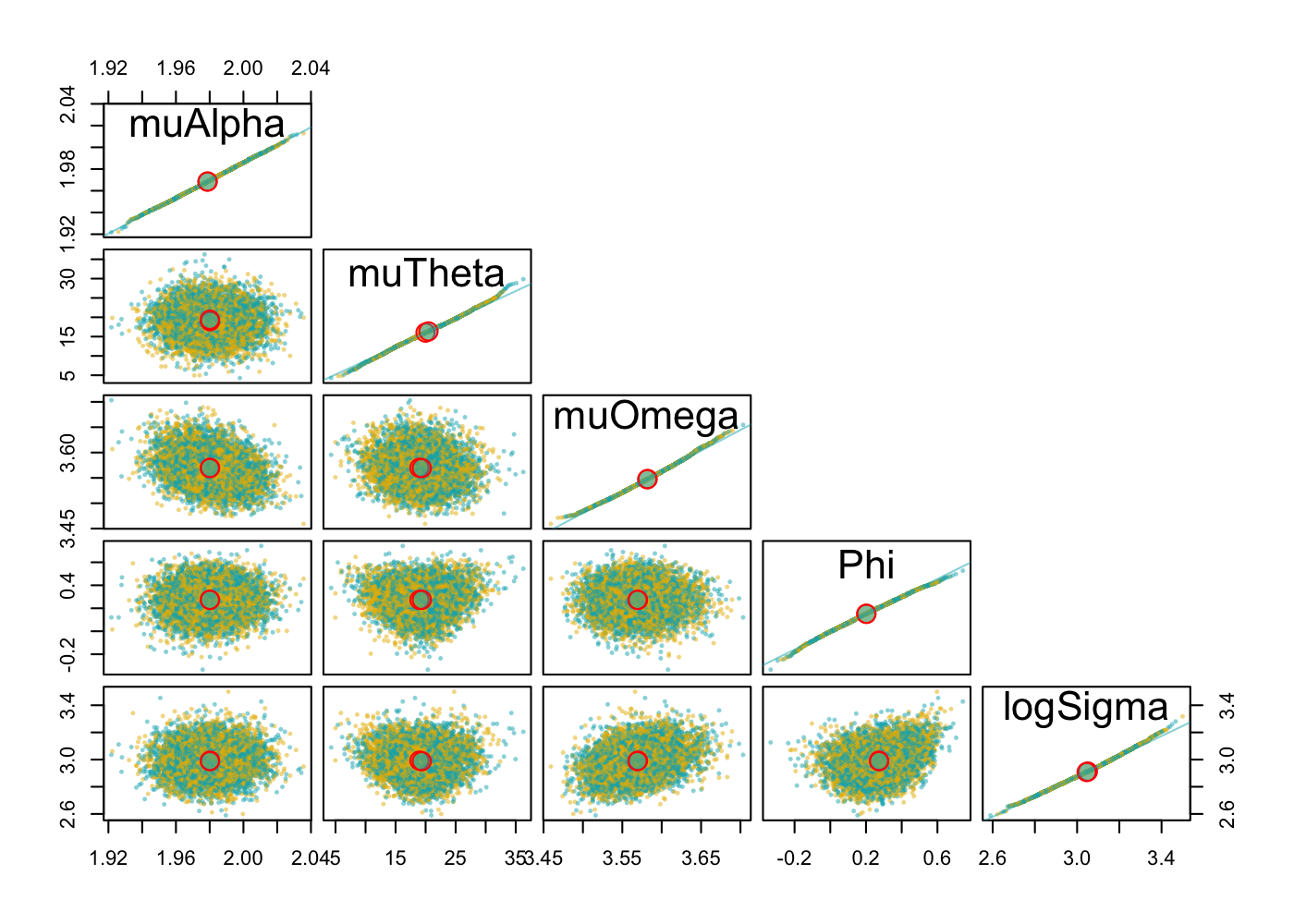}}}%
    \subfloat[$(tmax=50,n=100)$ with Prior2.]{{\includegraphics[width=8.5cm]{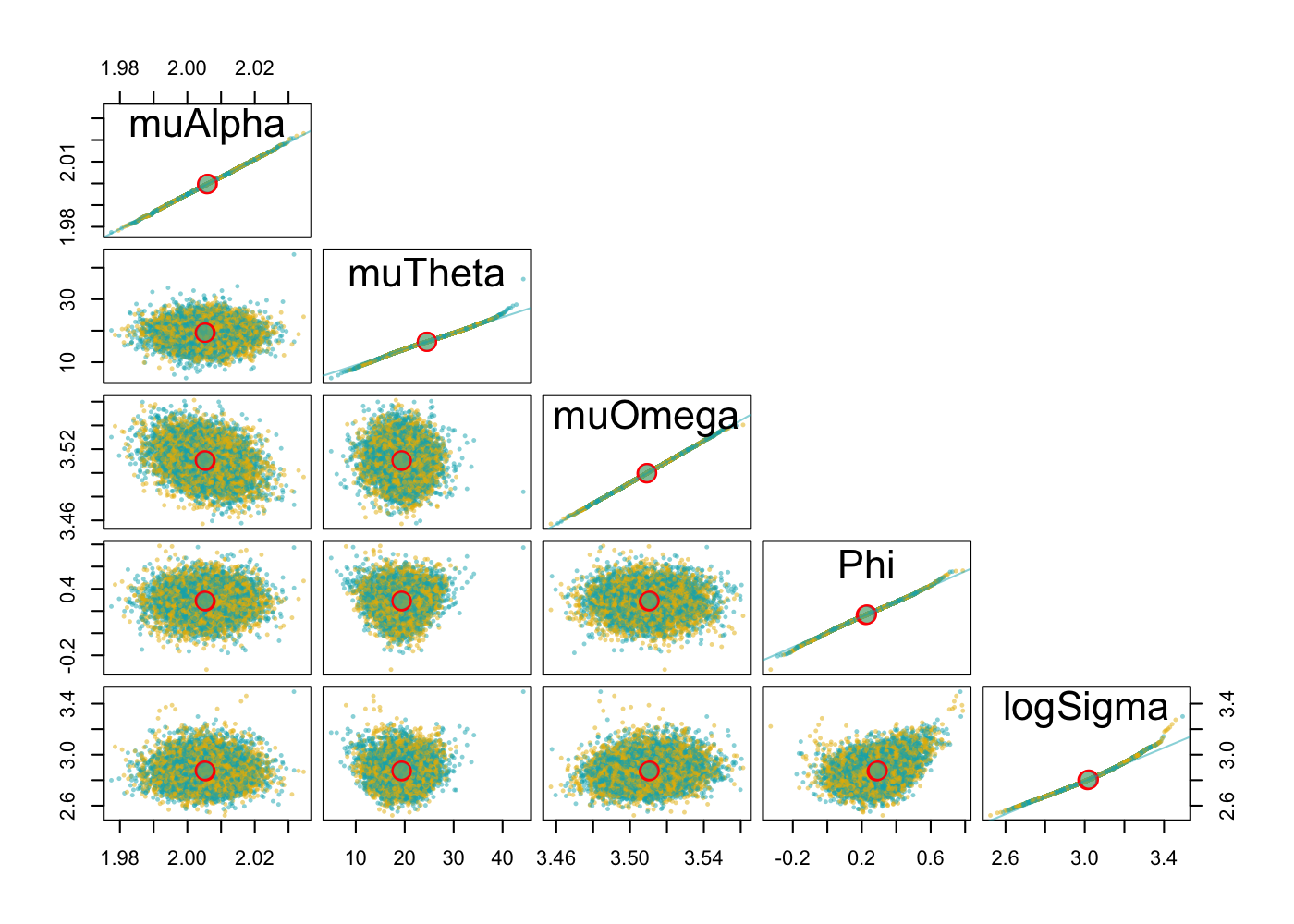}}}%
    \caption{Pair plots of posterior samples for Laplace approximation check from one realization of the simulation with Prior2. The four plots (a) (b) (c) and (d) correspond to the four settings of sample size in simulation. The random effects in the TMB model can be integrated with two techniques: (1) full MCMC integration via NUTS and  (2) Laplace approximation. To check the accuracy of Laplace approximation to the posterior likelihood density, the posterior samples for all the fixed effects in the model without (yellow dots) and with Laplace approximation (green dots) are shown pair-wisely on the same plot. Columns and rows on the lower diagonal correspond to pair-wise parameters, with the diagonal showing QQ-plot of posterior samples from Bayesian inference without (yellow dots) and with (green dots)
    Laplace approximation for that parameter including a 1:1 line in yellow. The large red circles on the off-diagonal plots represent the pairwise means. On each off-diagonal plot, there are 4000 yellow dots corresponding to 1000 samples retained from each of four chains without Laplace approximation, so as the green dots with Laplace approximation. Posterior rows were randomized to prevent consistent overplotting of one integration technique. Overlaps in the two colored dots suggest that the Laplace approximation is accurate.}
    \label{fig:lacheck}%
\end{figure}

By comparing the Bayesian posteriors with and without Laplace approximation, we are allowed to check how well the Laplace approximation works. Fig.~\ref{fig:lacheck} shows pair plots of posterior samples with and without Laplace approximation done by TMB under different sample size
settings with Prior2. Only autocorrelation in $\theta_t$ $\phi_{\theta,\theta}=0.4$ was considered.  Plot (a), (b), (c) and (d) correspond to
setting $(tmax=25, n=50)$, $(tmax=25, n=100)$, $(tmax=50, n=25)$ and $(tmax=50, n=100)$ respectively. On each subplot, the lower diagonal plots contain pairwise parameter posterior points. The green dots represent posterior points from full MCMC integration via NUTS and the yellow points from enabled Laplace approximation of the random effects.  The hollow red circles on the off-diagonal plots represent the pairwise means. The diagonal shows QQ-plot of posterior samples from Bayesian inference without (yellow dots) and with (green dots) Laplace approximation for that parameter including a 1:1 line in yellow. Even though the posterior points are densely packed, the overlap of the red circles with each technique shows seemingly good alignment of the two versions of the posterior, and this suggests that the Laplace approximation to the marginal likelihood where random effects are integrated out works well. Similar pair plots for Laplace approximation check with Prior1 can be found in Supporting Information.

\section{Real-data case study}
Having established the utility of our modeling approach and
frequentist and Bayesian inference in the context of simulated data, we also applied the same statistical model to the analysis of a real great tit dataset of practical interest. 
The observed data were collected from a Dutch great tit (\textit{Parus major}) population at
the Hoge Veluwe National Park in the Netherlands (52°02' - 52°07'N, 5°51' - 5°32E).
The recorded variables include the number of chicks, number of fledglings, mother ID, brood laying date and so on for each brood. Laying dates are presented as the
number of days after March 31 (day 1=April 1, day 31=May 1). Similar to \shortciteA{Reed2013}, 
only the broods with one or more chicks were considered in our analysis due to the high proportion (15.7\%)
of zero-observations in the number of fledglings among the broods. The number of fledglings was
taken as the fitness component and assumed to be Poisson distributed. The analyzed dataset consists of  brood records breeding in
61 years from 1955 to 2015 and the sample size in a specific year ranges from 10 to 164 with an average of 81 across the study years. See \citeA{Reed2013} for more details on the study population and fieldwork procedures.
 
The focus of this empirical study is to compare the computational efficiency of Bayesian inference with 
and without Laplace approximation and to check the accuracy of Laplace approximation.
However, since the true structure of the model is unknown, we first conducted model selection under the frequentist framework and the candidate models considered are different from each other only in the model structure of stochastic $\alpha_t$, $\theta_t$ and $\omega_t$. The details of all the candidate models including the best model are given in Supporting Information. 
We then made Bayesian inference with the two different priors as in the simulation study using the selected model. For each prior distribution,
we implemented tmbstan with and without Laplace approximation to check the accuracy of Laplace approximation.
 
Table \ref{table:2} lists the reported estimates of model parameters from maximum likelihood (MLE)
and Bayesian estimates with half-Cauchy (0, 10) prior (Prior1) and log-normal (1, 0.5) prior (Prior2). The best
model indicates VAR(1) structure of $\alpha_t$ and $\theta_t$ and non-zero correlation $\hat{\rho}_{\alpha,\theta}$.
The width of stabilizing fitness function turned to be constant over the study years implied by zero $\hat{\omega}_t$. Frequentist inference and Bayesian inference with Prior2 report close estimates for $\phi_{\theta,\theta}$ but the estimates
with Prior2 show again less uncertainty for most of the estimates except for $\rho_{\alpha,\theta}$. In terms of $\log\sigma_\theta$, 
Bayesian inference with Prior1 reports the largest estimate and least certainty compared with the other two techniques. The close resemblance between estimates of $\log\sigma_\theta$ based on
maximum likelihood and Bayesian inferences suggests that the data contains a good amount of information on $\log\sigma_\theta$
so that the maximum likelihood overwhelms the log-normal prior and dominates the posterior likelihood.
 
\begin{table}[]
\caption{Frequentist and Bayesian estimates of parameters in the selected model with great tit dataset. The
Bayesian estimates (in column Prior1 and Prior2) are obtained without Lapalace approximation done by TMB. }
\label{table:2}
\centering
\begin{tabular}{|l|l|l|l|l|l|l|}
\hline
\textbf{parameter}  & \textbf{MLE} & \textbf{Prior1} & \textbf{Prior2}\\ \hline
  $\mu_\alpha$ &2(0.0369)&2(0.0491)&2(0.0379) \\ \hline
  $\mu_\theta$ &18.5(5.35)&18.8(7.12)&19.4(5.09) \\ \hline
  $\mu_\omega$ &3.88(0.055)&3.89(0.0563)&3.86(0.0522) \\ \hline
  $\phi_{\alpha,\alpha}$ &0.379(0.12)&0.458(0.13)&0.398(0.124) \\ \hline
  $\phi_{\theta,\theta}$ &0.48(0.112)&0.545(0.114)&0.477(0.102) \\ \hline
  $log\sigma_\alpha$ &-1.72(0.14)&-1.63(0.152)&-1.76(0.126) \\ \hline
  $log\sigma_\theta$ &3.07(0.137)&3.16(0.155)&2.98(0.125) \\ \hline
  $\rho_{\alpha,\theta}$ &-0.728(0.0825)&-0.715(0.0895)&-0.661(0.0987) \\ \hline
\end{tabular}
\end{table}

 Table \ref{table:3} shows computational efficiencies of Bayesian inference without and with Laplace approximation. It turns out that the computational efficiency with Laplace approximation is approximately
half of that without Laplace approximation in both models with Prior1 and Prior2. 

\begin{table}[]
\centering
\caption{Comparison of computational efficiency between Bayesian inference without (in the row "Full MCMC") and with
Laplace approximation (in the row "Laplace approximation") for random effects for the great tit case study.}
\label{table:3}
\begin{tabular}{|l|l|l|l|l|}
\hline
\textbf{Model}           & \textbf{Inference}    & \textbf{Time(s)} & \textbf{min.ESS} & \textbf{Efficiency(ESS/t)} \\ \hline
\multirow{2}{*}{Prior 1} & Full MCMC         & 1542.215       &  186.7651      &  0.1211019              \\ \cline{2-5} 
                         & Laplace approximation  & 15491.85     & 1004.643     &  0.06484975                    \\ \hline
\multirow{2}{*}{Prior 2} & Full MCMC         &  1266.096      &  291.0717       & 0.229897               \\ \cline{2-5} 
                         & Laplace approximation  & 7815.218     & 1111.257   &  0.1421914                     \\ \hline
\end{tabular}
\end{table}

\begin{figure}[h!]
\centering
\includegraphics[scale=0.34]{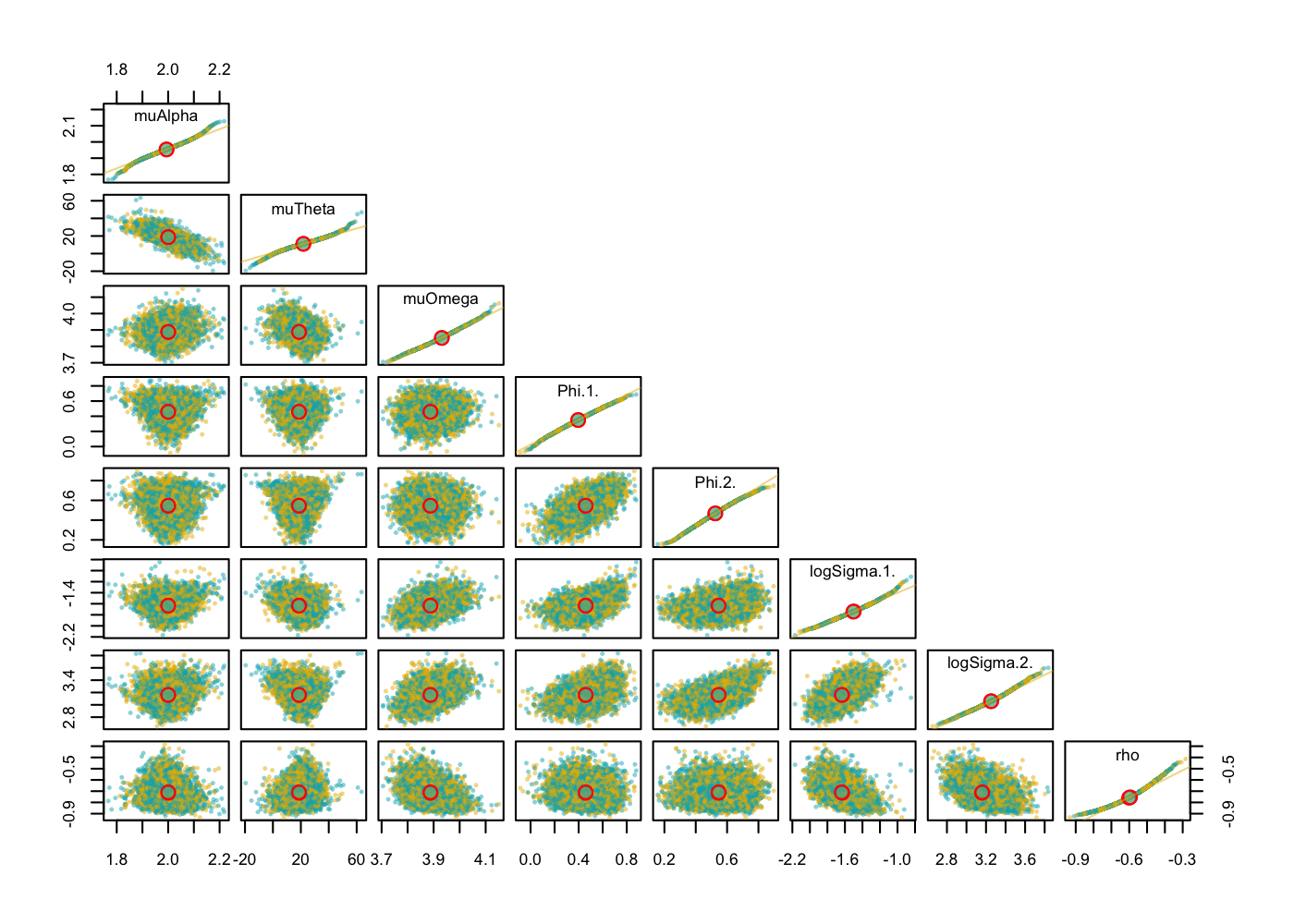}
\caption{Pair plots of posterior samples for Laplace approximation test for the great tit case study with Prior1.
The random effects in the great tit TMB model can be integrated with two techniques: (1) full MCMC integration via NUTS and 
(2) Laplace approximation. To check the accuracy of Laplace approximation to the posterior likelihood density, the posterior
samples for all the fixed effects in the model without (yellow dots) and with Laplace approximation (green dots) are shown pair-wisely on the same plot. Columns and rows on the lower diagonal correspond to pair-wise parameters, with the diagonal showing QQ-plot of posterior samples from Bayesian inference without (yellow dots) and with (green dots) Laplace approximation for that parameter including a 1:1 line in yellow. 
The large red circles of the off-diagonal plots represent the pairwise means. On each off-diagonal plot, there are 4000 yellow dots
corresponding to 1000 samples retained from each of four chains without Laplace approximation, so as the green dots with Laplace approximation. Posterior rows were randomized to prevent consistent overplotting of one integration technique. Overlaps in the two colored dots suggest the Laplace approximation assumption is met.}
\label{fig:gtp1}
\end{figure}

\begin{figure}[h!]
\centering
\includegraphics[scale=0.34]{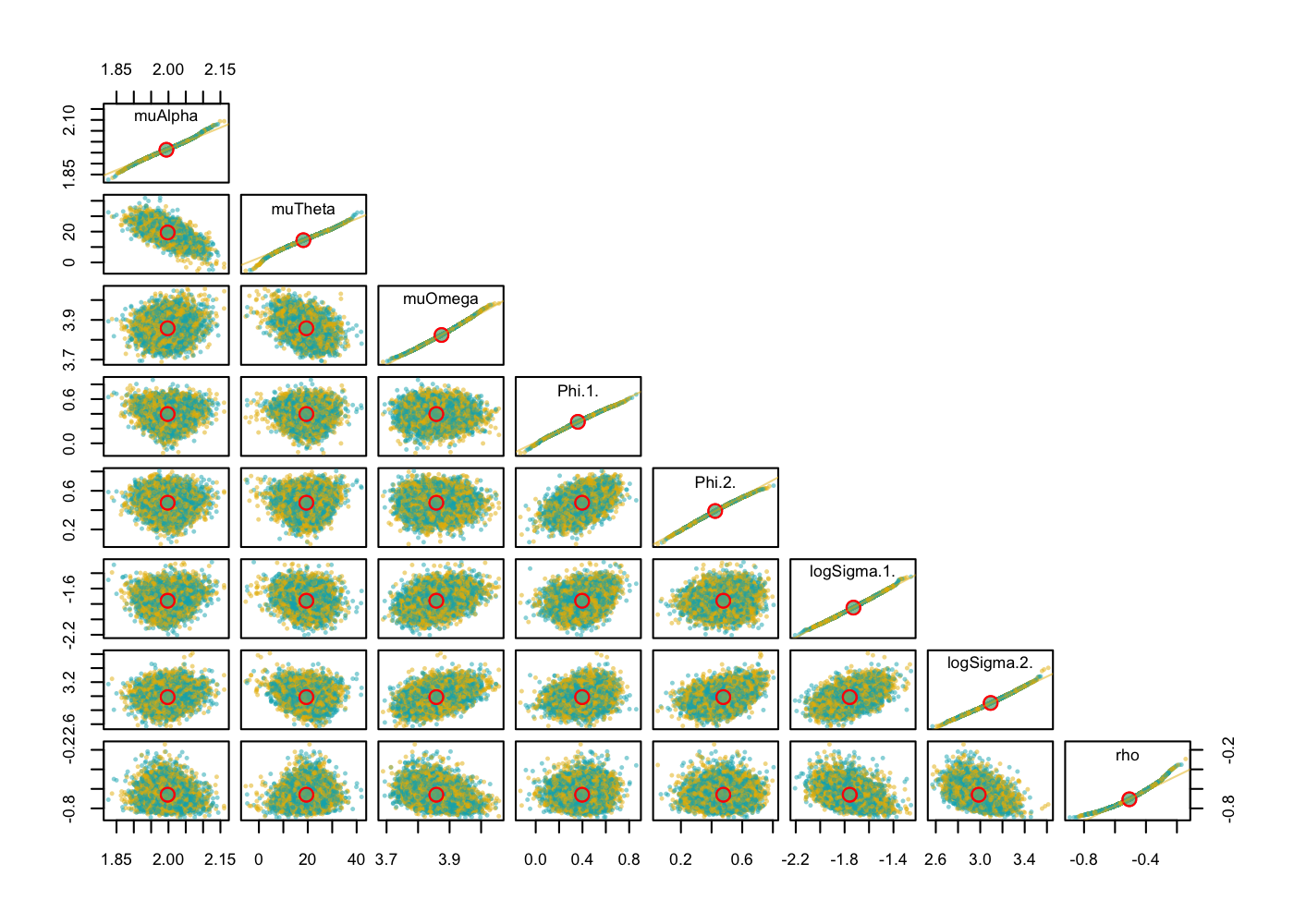}
\caption{Pair plots of posterior samples for Laplace approximation test for the great tit case study with Prior2. 
The random effects in the great tit TMB model can be integrated with two techniques: (1) full MCMC integration via NUTS and 
(2) Laplace approximation. To check the accuracy of Laplace approximation to the posterior likelihood density, the posterior
samples for all the fixed effects in the model without (yellow dots) and with Laplace approximation (green dots) are shown pair-wisely on the same plot. Columns and rows on the lower diagonal correspond to pair-wise parameters, with the diagonal showing QQ-plot of posterior samples from Bayesian inference without (yellow dots) and with (green dots) Laplace approximation for that parameter including a 1:1 line in yellow. 
The large red circles of the off-diagonal plots represent the pairwise means. On each off-diagonal plot, 4000 yellow dots correspond to 1000 samples retained from each of four chains without Laplace approximation, so as the green dots with Laplace approximation. Posterior rows were randomized to prevent consistent overplotting of one integration technique. Overlaps in the two colored dots suggest the Laplace approximation assumption is met.}
\label{fig:gtp2}
\end{figure}

Similar to Fig.~\ref{fig:lacheck}, Fig.~\ref{fig:gtp1} and Fig.~\ref{fig:gtp2} display pair plots of posterior samples to check the accuracy of Laplace approximation with Prior1 and Prior2 respectively. Both the figures seemingly suggest a good mix of posterior samples with and without Laplace
approximation for all the parameters in the selected model, indicating that the Laplace approximation assumption is met.

\section{Conclusions and extensions}
In this study, we have investigated frequentist inference and Bayesian inference with two different priors. The inferences were
implemented with a state-space model estimating temporal fluctuating selection and with simulated biological data under four different simulation settings. A state-of-the-art R package (tmbstan) for fast fitting statistical models was used for Bayesian inference with Laplace approximation turning on or off. The simulation studies show that the choice of prior can have an important impact on the geometric shape of the posterior distributions of the model parameters and a non-informative prior (in this study uniform prior and inverse-gamma prior on the scale parameter) may lead to unstable inference since the Markov chains may not converge or get stuck in part of the ridge of posterior. With unobserved states following a VAR(1) process, we also found that the autoregressive parameters and the scale parameters in the variance-covariance matrix of the states are difficult and challenging to be estimated accurately. The increased sample size at each time point does not necessarily provide more information for the transition parameters and scale parameters. Only more time points in the data could make the likelihood dominate the posterior likelihood and thus lead to better estimates of these parameters.  Half-Cauchy prior on the scale parameter leads to less stable inference than log-normal prior indicated by the number of divergent transitions in the Markov Chains. Laplace approximation for the random effects turns out to be accurate suggested by the pair plots of the posterior samples with and without Laplace approximation for both the simulation studies and the great tit case study. Turning on Laplace approximation in tmbstan would probably reduce computational efficiency but it is worth trying when there is a good amount of data, in which case the Laplace approximation is more likely to be accurate and also potentially improve the computational efficiency of MCMC.

In our study, we used arbitrary prior distributions, however, the prior information can be obtained from different sources. For example, in our great tit case study, the timing and width of the caterpillar peak can provide a clue for the time window of optimal laying dates, thus the information can be used to decide the prior for the scale parameter of the optimal laying dates. Prior information can also be generated from previous studies on the same species and more general ecological knowledge coming from other related species \cite{tufto2000bayesian}.

We conducted simulation studies with only AR(1) process of the optimal laying dates, but the model is formulated and coded in a way that can be effortlessly extended to order-1 vector autoregression (VAR(1)). It can be widely used for modeling ecological processes where auto-correlation and cross-correlation in the processes arise due to shared environmental variables at either temporal or spatial scale. We expect more ecologists to adopt these two new estimation methods, TMB, and tmbstan, given its flexibility in either frequentist or Bayesian inference for a wide range of models, including the models where the unobserved ecological processes are treated as latent variables and assumed to be VAR processes. However, the drawback of Bayesian VAR (BVAR) methods is that it usually requires estimation of a large number of parameters and thus the over-parameterization might lead to unstable inference and inaccurate out-of-sample forecasts. Some shrinkage methods \shortcite[for example]{sims1998bayesian, koop2010bayesian, giannone2015prior,sorbye2017penalised} were thereby developed, in which Bayesian priors provide a logical and consistent method of imposing parameter restrictions that can be potentially applied to ecological data cases.

\bibliographystyle{apacite}

\newpage
\begin{thebibliography}{}

\bibitem [\protect \citeauthoryear {%
Albertsen%
, Nielsen%
\BCBL {}\ \BBA {} Thygesen%
}{%
Albertsen%
\ \protect \BOthers {.}}{%
{\protect \APACyear {2016}}%
}]{%
albertsen2016choosing}
\APACinsertmetastar {%
albertsen2016choosing}%
\begin{APACrefauthors}%
Albertsen, C\BPBI M.%
, Nielsen, A.%
\BCBL {}\ \BBA {} Thygesen, U\BPBI H.%
\end{APACrefauthors}%
\unskip\
\newblock
\APACrefYearMonthDay{2016}{}{}.
\newblock
{\BBOQ}\APACrefatitle {Choosing the observational likelihood in state-space
  stock assessment models} {Choosing the observational likelihood in
  state-space stock assessment models}.{\BBCQ}
\newblock
\APACjournalVolNumPages{Canadian Journal of Fisheries and Aquatic
  Sciences}{74}{5}{779--789}.
\newblock
\begin{APACrefDOI} \doi{10.1139/cjfas-2015-0532} \end{APACrefDOI}
\PrintBackRefs{\CurrentBib}

\bibitem [\protect \citeauthoryear {%
Auger-M{\'e}th{\'e}%
\ \protect \BOthers {.}}{%
Auger-M{\'e}th{\'e}%
\ \protect \BOthers {.}}{%
{\protect \APACyear {2017}}%
}]{%
auger2017spatiotemporal}
\APACinsertmetastar {%
auger2017spatiotemporal}%
\begin{APACrefauthors}%
Auger-M{\'e}th{\'e}, M.%
, Albertsen, C\BPBI M.%
, Jonsen, I\BPBI D.%
, Derocher, A\BPBI E.%
, Lidgard, D\BPBI C.%
, Studholme, K\BPBI R.%
\BDBL {}Flemming, J\BPBI M.%
\end{APACrefauthors}%
\unskip\
\newblock
\APACrefYearMonthDay{2017}{}{}.
\newblock
{\BBOQ}\APACrefatitle {Spatiotemporal modelling of marine movement data using
  {Template Model Builder (TMB)}} {Spatiotemporal modelling of marine movement
  data using {Template Model Builder (TMB)}}.{\BBCQ}
\newblock
\APACjournalVolNumPages{Marine Ecology Progress Series}{565}{}{237--249}.
\newblock
\begin{APACrefDOI} \doi{10.3354/meps12019} \end{APACrefDOI}
\PrintBackRefs{\CurrentBib}

\bibitem [\protect \citeauthoryear {%
Betancourt%
\ \BBA {} Girolami%
}{%
Betancourt%
\ \BBA {} Girolami%
}{%
{\protect \APACyear {2015}}%
}]{%
betancourt2015hamiltonian}
\APACinsertmetastar {%
betancourt2015hamiltonian}%
\begin{APACrefauthors}%
Betancourt, M.%
\BCBT {}\ \BBA {} Girolami, M.%
\end{APACrefauthors}%
\unskip\
\newblock
\APACrefYearMonthDay{2015}{}{}.
\newblock
{\BBOQ}\APACrefatitle {{Hamiltonian Monte Carlo} for hierarchical models}
  {{Hamiltonian Monte Carlo} for hierarchical models}.{\BBCQ}
\newblock
\APACjournalVolNumPages{Current trends in Bayesian methodology with
  applications}{79}{}{30}.
\PrintBackRefs{\CurrentBib}

\bibitem [\protect \citeauthoryear {%
Cadigan%
}{%
Cadigan%
}{%
{\protect \APACyear {2015}}%
}]{%
cadigan2015state}
\APACinsertmetastar {%
cadigan2015state}%
\begin{APACrefauthors}%
Cadigan, N\BPBI G.%
\end{APACrefauthors}%
\unskip\
\newblock
\APACrefYearMonthDay{2015}{}{}.
\newblock
{\BBOQ}\APACrefatitle {A state-space stock assessment model for northern cod,
  including under-reported catches and variable natural mortality rates} {A
  state-space stock assessment model for northern cod, including under-reported
  catches and variable natural mortality rates}.{\BBCQ}
\newblock
\APACjournalVolNumPages{Canadian Journal of Fisheries and Aquatic
  Sciences}{73}{2}{296--308}.
\newblock
\begin{APACrefDOI} \doi{10.1139/cjfas-2015-0047} \end{APACrefDOI}
\PrintBackRefs{\CurrentBib}

\bibitem [\protect \citeauthoryear {%
Cao%
, Visser%
\BCBL {}\ \BBA {} Tufto%
}{%
Cao%
\ \protect \BOthers {.}}{%
{\protect \APACyear {2019}}%
}]{%
cao2019time}
\APACinsertmetastar {%
cao2019time}%
\begin{APACrefauthors}%
Cao, Y.%
, Visser, M\BPBI E.%
\BCBL {}\ \BBA {} Tufto, J.%
\end{APACrefauthors}%
\unskip\
\newblock
\APACrefYearMonthDay{2019}{}{}.
\newblock
{\BBOQ}\APACrefatitle {A time-series model for estimating temporal variation in
  phenotypic selection on laying dates in a Dutch great tit population} {A
  time-series model for estimating temporal variation in phenotypic selection
  on laying dates in a dutch great tit population}.{\BBCQ}
\newblock
\APACjournalVolNumPages{Methods in Ecology and Evolution}{10}{9}{1401--1411}.
\PrintBackRefs{\CurrentBib}

\bibitem [\protect \citeauthoryear {%
Carpenter%
\ \protect \BOthers {.}}{%
Carpenter%
\ \protect \BOthers {.}}{%
{\protect \APACyear {2017}}%
}]{%
carpenter2017stan}
\APACinsertmetastar {%
carpenter2017stan}%
\begin{APACrefauthors}%
Carpenter, B.%
, Gelman, A.%
, Hoffman, M\BPBI D.%
, Lee, D.%
, Goodrich, B.%
, Betancourt, M.%
\BDBL {}Riddell, A.%
\end{APACrefauthors}%
\unskip\
\newblock
\APACrefYearMonthDay{2017}{}{}.
\newblock
{\BBOQ}\APACrefatitle {Stan: A probabilistic programming language} {Stan: A
  probabilistic programming language}.{\BBCQ}
\newblock
\APACjournalVolNumPages{Journal of statistical software}{76}{1}{}.
\PrintBackRefs{\CurrentBib}

\bibitem [\protect \citeauthoryear {%
Chevin%
, Visser%
\BCBL {}\ \BBA {} Tufto%
}{%
Chevin%
\ \protect \BOthers {.}}{%
{\protect \APACyear {2015}}%
}]{%
Chevin2015}
\APACinsertmetastar {%
Chevin2015}%
\begin{APACrefauthors}%
Chevin, L\BHBI M.%
, Visser, M\BPBI E.%
\BCBL {}\ \BBA {} Tufto, J.%
\end{APACrefauthors}%
\unskip\
\newblock
\APACrefYearMonthDay{2015}{}{}.
\newblock
{\BBOQ}\APACrefatitle {{Estimating the variation, autocorrelation, and
  environmental sensitivity of phenotypic selection}} {{Estimating the
  variation, autocorrelation, and environmental sensitivity of phenotypic
  selection}}.{\BBCQ}
\newblock
\APACjournalVolNumPages{Evolution}{69}{9}{2319--2332}.
\newblock
\begin{APACrefDOI} \doi{10.1111/evo.12741} \end{APACrefDOI}
\PrintBackRefs{\CurrentBib}

\bibitem [\protect \citeauthoryear {%
Fuglstad%
, Hem%
, Knight%
, Rue%
\BCBL {}\ \BBA {} Riebler%
}{%
Fuglstad%
\ \protect \BOthers {.}}{%
{\protect \APACyear {2019}}%
}]{%
fuglstad2019intuitive}
\APACinsertmetastar {%
fuglstad2019intuitive}%
\begin{APACrefauthors}%
Fuglstad, G\BHBI A.%
, Hem, I\BPBI G.%
, Knight, A.%
, Rue, H.%
\BCBL {}\ \BBA {} Riebler, A.%
\end{APACrefauthors}%
\unskip\
\newblock
\APACrefYearMonthDay{2019}{}{}.
\newblock
{\BBOQ}\APACrefatitle {Intuitive principle-based priors for attributing
  variance in additive model structures} {Intuitive principle-based priors for
  attributing variance in additive model structures}.{\BBCQ}
\newblock
\APACjournalVolNumPages{arXiv preprint arXiv:1902.00242}{}{}{}.
\PrintBackRefs{\CurrentBib}

\bibitem [\protect \citeauthoryear {%
Gelman%
, Lee%
\BCBL {}\ \BBA {} Guo%
}{%
Gelman%
\ \protect \BOthers {.}}{%
{\protect \APACyear {2015}}%
}]{%
gelman2015stan}
\APACinsertmetastar {%
gelman2015stan}%
\begin{APACrefauthors}%
Gelman, A.%
, Lee, D.%
\BCBL {}\ \BBA {} Guo, J.%
\end{APACrefauthors}%
\unskip\
\newblock
\APACrefYearMonthDay{2015}{}{}.
\newblock
{\BBOQ}\APACrefatitle {Stan: A probabilistic programming language for Bayesian
  inference and optimization} {Stan: A probabilistic programming language for
  bayesian inference and optimization}.{\BBCQ}
\newblock
\APACjournalVolNumPages{Journal of Educational and Behavioral
  Statistics}{40}{5}{530--543}.
\PrintBackRefs{\CurrentBib}

\bibitem [\protect \citeauthoryear {%
Gelman%
\ \protect \BOthers {.}}{%
Gelman%
\ \protect \BOthers {.}}{%
{\protect \APACyear {2006}}%
}]{%
gelman2006prior}
\APACinsertmetastar {%
gelman2006prior}%
\begin{APACrefauthors}%
Gelman, A.%
\BCBT {}\ \BOthersPeriod {.}
\end{APACrefauthors}%
\unskip\
\newblock
\APACrefYearMonthDay{2006}{}{}.
\newblock
{\BBOQ}\APACrefatitle {Prior distributions for variance parameters in
  hierarchical models (comment on article by {Browne and Draper})} {Prior
  distributions for variance parameters in hierarchical models (comment on
  article by {Browne and Draper})}.{\BBCQ}
\newblock
\APACjournalVolNumPages{Bayesian analysis}{1}{3}{515--534}.
\PrintBackRefs{\CurrentBib}

\bibitem [\protect \citeauthoryear {%
Giannone%
, Lenza%
\BCBL {}\ \BBA {} Primiceri%
}{%
Giannone%
\ \protect \BOthers {.}}{%
{\protect \APACyear {2015}}%
}]{%
giannone2015prior}
\APACinsertmetastar {%
giannone2015prior}%
\begin{APACrefauthors}%
Giannone, D.%
, Lenza, M.%
\BCBL {}\ \BBA {} Primiceri, G\BPBI E.%
\end{APACrefauthors}%
\unskip\
\newblock
\APACrefYearMonthDay{2015}{}{}.
\newblock
{\BBOQ}\APACrefatitle {Prior selection for vector autoregressions} {Prior
  selection for vector autoregressions}.{\BBCQ}
\newblock
\APACjournalVolNumPages{Review of Economics and Statistics}{97}{2}{436--451}.
\PrintBackRefs{\CurrentBib}

\bibitem [\protect \citeauthoryear {%
Grant%
\ \BBA {} Grant%
}{%
Grant%
\ \BBA {} Grant%
}{%
{\protect \APACyear {2002}}%
}]{%
grant2002unpredictable}
\APACinsertmetastar {%
grant2002unpredictable}%
\begin{APACrefauthors}%
Grant, P\BPBI R.%
\BCBT {}\ \BBA {} Grant, B\BPBI R.%
\end{APACrefauthors}%
\unskip\
\newblock
\APACrefYearMonthDay{2002}{}{}.
\newblock
{\BBOQ}\APACrefatitle {Unpredictable evolution in a 30-year study of {D}arwin's
  finches} {Unpredictable evolution in a 30-year study of {D}arwin's
  finches}.{\BBCQ}
\newblock
\APACjournalVolNumPages{science}{296}{5568}{707--711}.
\newblock
\begin{APACrefDOI} \doi{10.1126/science.1070315} \end{APACrefDOI}
\PrintBackRefs{\CurrentBib}

\bibitem [\protect \citeauthoryear {%
Hoffman%
\ \BBA {} Gelman%
}{%
Hoffman%
\ \BBA {} Gelman%
}{%
{\protect \APACyear {2014}}%
}]{%
hoffman2014no}
\APACinsertmetastar {%
hoffman2014no}%
\begin{APACrefauthors}%
Hoffman, M\BPBI D.%
\BCBT {}\ \BBA {} Gelman, A.%
\end{APACrefauthors}%
\unskip\
\newblock
\APACrefYearMonthDay{2014}{}{}.
\newblock
{\BBOQ}\APACrefatitle {The {No-U-Turn} sampler: adaptively setting path lengths
  in {Hamiltonian Monte Carlo}.} {The {No-U-Turn} sampler: adaptively setting
  path lengths in {Hamiltonian Monte Carlo}.}{\BBCQ}
\newblock
\APACjournalVolNumPages{Journal of Machine Learning
  Research}{15}{1}{1593--1623}.
\PrintBackRefs{\CurrentBib}

\bibitem [\protect \citeauthoryear {%
Jeffreys%
\ \BBA {} Jeffreys%
}{%
Jeffreys%
\ \BBA {} Jeffreys%
}{%
{\protect \APACyear {1961}}%
}]{%
jeffreys1961theory}
\APACinsertmetastar {%
jeffreys1961theory}%
\begin{APACrefauthors}%
Jeffreys, H.%
\BCBT {}\ \BBA {} Jeffreys, H.%
\end{APACrefauthors}%
\unskip\
\newblock
\APACrefYearMonthDay{1961}{}{}.
\newblock
\APACrefbtitle {Theory of Probability (3rd edn).} {Theory of probability (3rd
  edn).}
\newblock
\APACaddressPublisher{}{Oxford}.
\PrintBackRefs{\CurrentBib}

\bibitem [\protect \citeauthoryear {%
Karakani%
, van Niekerk%
\BCBL {}\ \BBA {} van Staden%
}{%
Karakani%
\ \protect \BOthers {.}}{%
{\protect \APACyear {2016}}%
}]{%
karakani2016bayesian}
\APACinsertmetastar {%
karakani2016bayesian}%
\begin{APACrefauthors}%
Karakani, H\BPBI M.%
, van Niekerk, J.%
\BCBL {}\ \BBA {} van Staden, P.%
\end{APACrefauthors}%
\unskip\
\newblock
\APACrefYearMonthDay{2016}{}{}.
\newblock
{\BBOQ}\APACrefatitle {Bayesian Analysis of {AR} (1) model} {Bayesian analysis
  of {AR} (1) model}.{\BBCQ}
\newblock
\APACjournalVolNumPages{arXiv preprint arXiv:1611.08747}{}{}{}.
\PrintBackRefs{\CurrentBib}

\bibitem [\protect \citeauthoryear {%
Koop%
, Korobilis%
\BCBL {}\ \protect \BOthers {.}}{%
Koop%
\ \protect \BOthers {.}}{%
{\protect \APACyear {2010}}%
}]{%
koop2010bayesian}
\APACinsertmetastar {%
koop2010bayesian}%
\begin{APACrefauthors}%
Koop, G.%
, Korobilis, D.%
\BCBL {}\ \BOthersPeriod {.}\end{APACrefauthors}%
\unskip\
\newblock
\APACrefYearMonthDay{2010}{}{}.
\newblock
{\BBOQ}\APACrefatitle {Bayesian multivariate time series methods for empirical
  macroeconomics} {Bayesian multivariate time series methods for empirical
  macroeconomics}.{\BBCQ}
\newblock
\APACjournalVolNumPages{Foundations and Trends{\textregistered} in
  Econometrics}{3}{4}{267--358}.
\PrintBackRefs{\CurrentBib}

\bibitem [\protect \citeauthoryear {%
Kristensen%
, Nielsen%
, Berg%
, Skaug%
\BCBL {}\ \BBA {} Bell%
}{%
Kristensen%
\ \protect \BOthers {.}}{%
{\protect \APACyear {2016}}%
}]{%
kristensen2016tmb}
\APACinsertmetastar {%
kristensen2016tmb}%
\begin{APACrefauthors}%
Kristensen, K.%
, Nielsen, A.%
, Berg, C\BPBI W.%
, Skaug, H.%
\BCBL {}\ \BBA {} Bell, B\BPBI M.%
\end{APACrefauthors}%
\unskip\
\newblock
\APACrefYearMonthDay{2016}{}{}.
\newblock
{\BBOQ}\APACrefatitle {{TMB}: Automatic Differentiation and {L}aplace
  Approximation} {{TMB}: Automatic differentiation and {L}aplace
  approximation}.{\BBCQ}
\newblock
\APACjournalVolNumPages{Journal of Statistical Software}{70}{5}{1--21}.
\newblock
\begin{APACrefDOI} \doi{10.18637/jss.v070.i05} \end{APACrefDOI}
\PrintBackRefs{\CurrentBib}

\bibitem [\protect \citeauthoryear {%
Lei%
, Boys%
, Gillespie%
, Greenall%
\BCBL {}\ \BBA {} Wilkinson%
}{%
Lei%
\ \protect \BOthers {.}}{%
{\protect \APACyear {2011}}%
}]{%
lei2011bayesian}
\APACinsertmetastar {%
lei2011bayesian}%
\begin{APACrefauthors}%
Lei, G.%
, Boys, R.%
, Gillespie, C.%
, Greenall, A.%
\BCBL {}\ \BBA {} Wilkinson, D.%
\end{APACrefauthors}%
\unskip\
\newblock
\APACrefYearMonthDay{2011}{}{}.
\newblock
{\BBOQ}\APACrefatitle {Bayesian inference for sparse {VAR} (1) models, with
  application to time course microarray data} {Bayesian inference for sparse
  {VAR} (1) models, with application to time course microarray data}.{\BBCQ}
\newblock
\APACjournalVolNumPages{Journal of Biometrics and Biostatistics}{}{}{}.
\PrintBackRefs{\CurrentBib}

\bibitem [\protect \citeauthoryear {%
Lunn%
, Thomas%
, Best%
\BCBL {}\ \BBA {} Spiegelhalter%
}{%
Lunn%
\ \protect \BOthers {.}}{%
{\protect \APACyear {2000}}%
}]{%
lunn2000winbugs}
\APACinsertmetastar {%
lunn2000winbugs}%
\begin{APACrefauthors}%
Lunn, D\BPBI J.%
, Thomas, A.%
, Best, N.%
\BCBL {}\ \BBA {} Spiegelhalter, D.%
\end{APACrefauthors}%
\unskip\
\newblock
\APACrefYearMonthDay{2000}{}{}.
\newblock
{\BBOQ}\APACrefatitle {{WinBUGS}-a {Bayesian} modelling framework: concepts,
  structure, and extensibility} {{WinBUGS}-a {Bayesian} modelling framework:
  concepts, structure, and extensibility}.{\BBCQ}
\newblock
\APACjournalVolNumPages{Statistics and computing}{10}{4}{325--337}.
\PrintBackRefs{\CurrentBib}

\bibitem [\protect \citeauthoryear {%
Monnahan%
\ \BBA {} Kristensen%
}{%
Monnahan%
\ \BBA {} Kristensen%
}{%
{\protect \APACyear {2018}}%
}]{%
monnahan2018no}
\APACinsertmetastar {%
monnahan2018no}%
\begin{APACrefauthors}%
Monnahan, C\BPBI C.%
\BCBT {}\ \BBA {} Kristensen, K.%
\end{APACrefauthors}%
\unskip\
\newblock
\APACrefYearMonthDay{2018}{}{}.
\newblock
{\BBOQ}\APACrefatitle {{No-U-turn} sampling for fast Bayesian inference in
  {ADMB} and {TMB}: {I}ntroducing the adnuts and tmbstan {R} packages}
  {{No-U-turn} sampling for fast bayesian inference in {ADMB} and {TMB}:
  {I}ntroducing the adnuts and tmbstan {R} packages}.{\BBCQ}
\newblock
\APACjournalVolNumPages{PloS one}{13}{5}{e0197954}.
\PrintBackRefs{\CurrentBib}

\bibitem [\protect \citeauthoryear {%
Monnahan%
, Thorson%
\BCBL {}\ \BBA {} Branch%
}{%
Monnahan%
\ \protect \BOthers {.}}{%
{\protect \APACyear {2017}}%
}]{%
monnahan2017faster}
\APACinsertmetastar {%
monnahan2017faster}%
\begin{APACrefauthors}%
Monnahan, C\BPBI C.%
, Thorson, J\BPBI T.%
\BCBL {}\ \BBA {} Branch, T\BPBI A.%
\end{APACrefauthors}%
\unskip\
\newblock
\APACrefYearMonthDay{2017}{}{}.
\newblock
{\BBOQ}\APACrefatitle {Faster estimation of Bayesian models in ecology using
  {Hamiltonian Monte Carlo}} {Faster estimation of bayesian models in ecology
  using {Hamiltonian Monte Carlo}}.{\BBCQ}
\newblock
\APACjournalVolNumPages{Methods in Ecology and Evolution}{8}{3}{339--348}.
\PrintBackRefs{\CurrentBib}

\bibitem [\protect \citeauthoryear {%
Neal%
\ \protect \BOthers {.}}{%
Neal%
\ \protect \BOthers {.}}{%
{\protect \APACyear {2011}}%
}]{%
neal2011mcmc}
\APACinsertmetastar {%
neal2011mcmc}%
\begin{APACrefauthors}%
Neal, R\BPBI M.%
\BCBT {}\ \BOthersPeriod {.}
\end{APACrefauthors}%
\unskip\
\newblock
\APACrefYearMonthDay{2011}{}{}.
\newblock
{\BBOQ}\APACrefatitle {{MCMC} using {H}amiltonian dynamics} {{MCMC} using
  {H}amiltonian dynamics}.{\BBCQ}
\newblock
\APACjournalVolNumPages{Handbook of markov chain monte carlo}{2}{11}{2}.
\PrintBackRefs{\CurrentBib}

\bibitem [\protect \citeauthoryear {%
Reed%
, Jenouvrier%
\BCBL {}\ \BBA {} Visser%
}{%
Reed%
\ \protect \BOthers {.}}{%
{\protect \APACyear {2013}}%
}]{%
Reed2013}
\APACinsertmetastar {%
Reed2013}%
\begin{APACrefauthors}%
Reed, T\BPBI E.%
, Jenouvrier, S.%
\BCBL {}\ \BBA {} Visser, M\BPBI E.%
\end{APACrefauthors}%
\unskip\
\newblock
\APACrefYearMonthDay{2013}{}{}.
\newblock
{\BBOQ}\APACrefatitle {{Phenological mismatch strongly affects individual
  fitness but not population demography in a woodland passerine}}
  {{Phenological mismatch strongly affects individual fitness but not
  population demography in a woodland passerine}}.{\BBCQ}
\newblock
\APACjournalVolNumPages{Journal of Animal Ecology}{82}{1}{131--144}.
\newblock
\begin{APACrefDOI} \doi{10.1111/j.1365-2656.2012.02020.x} \end{APACrefDOI}
\PrintBackRefs{\CurrentBib}

\bibitem [\protect \citeauthoryear {%
Schlaifer%
\ \BBA {} Raiffa%
}{%
Schlaifer%
\ \BBA {} Raiffa%
}{%
{\protect \APACyear {1961}}%
}]{%
raiffa1961applied}
\APACinsertmetastar {%
raiffa1961applied}%
\begin{APACrefauthors}%
Schlaifer, R.%
\BCBT {}\ \BBA {} Raiffa, H.%
\end{APACrefauthors}%
\unskip\
\newblock
\APACrefYear{1961}.
\newblock
\APACrefbtitle {Applied statistical decision theory} {Applied statistical
  decision theory}.
\newblock
\APACaddressPublisher{}{Wiley Cambridge}.
\PrintBackRefs{\CurrentBib}

\bibitem [\protect \citeauthoryear {%
Sims%
\ \BBA {} Zha%
}{%
Sims%
\ \BBA {} Zha%
}{%
{\protect \APACyear {1998}}%
}]{%
sims1998bayesian}
\APACinsertmetastar {%
sims1998bayesian}%
\begin{APACrefauthors}%
Sims, C\BPBI A.%
\BCBT {}\ \BBA {} Zha, T.%
\end{APACrefauthors}%
\unskip\
\newblock
\APACrefYearMonthDay{1998}{}{}.
\newblock
{\BBOQ}\APACrefatitle {Bayesian methods for dynamic multivariate models}
  {Bayesian methods for dynamic multivariate models}.{\BBCQ}
\newblock
\APACjournalVolNumPages{International Economic Review}{}{}{949--968}.
\PrintBackRefs{\CurrentBib}

\bibitem [\protect \citeauthoryear {%
S{\o}rbye%
\ \BBA {} Rue%
}{%
S{\o}rbye%
\ \BBA {} Rue%
}{%
{\protect \APACyear {2017}}%
}]{%
sorbye2017penalised}
\APACinsertmetastar {%
sorbye2017penalised}%
\begin{APACrefauthors}%
S{\o}rbye, S\BPBI H.%
\BCBT {}\ \BBA {} Rue, H.%
\end{APACrefauthors}%
\unskip\
\newblock
\APACrefYearMonthDay{2017}{}{}.
\newblock
{\BBOQ}\APACrefatitle {Penalised complexity priors for stationary
  autoregressive processes} {Penalised complexity priors for stationary
  autoregressive processes}.{\BBCQ}
\newblock
\APACjournalVolNumPages{Journal of Time Series Analysis}{38}{6}{923--935}.
\PrintBackRefs{\CurrentBib}

\bibitem [\protect \citeauthoryear {%
{Stan Development Team}%
}{%
{Stan Development Team}%
}{%
{\protect \APACyear {2018}}%
{\protect \APACexlab {{\protect \BCnt {1}}}}}]{%
team2018rstan}
\APACinsertmetastar {%
team2018rstan}%
\begin{APACrefauthors}%
{Stan Development Team}.%
\end{APACrefauthors}%
\unskip\
\newblock
\APACrefYearMonthDay{2018{\protect \BCnt {1}}}{}{}.
\newblock
{\BBOQ}\APACrefatitle {RStan: the {R} interface to Stan} {Rstan: the {R}
  interface to stan}.{\BBCQ}
\newblock
\APACjournalVolNumPages{R package version 2.17.3}{}{}{}.
\newblock
\APAChowpublished {\url{http://mc-stan.org}}.
\PrintBackRefs{\CurrentBib}

\bibitem [\protect \citeauthoryear {%
{Stan Development Team}%
}{%
{Stan Development Team}%
}{%
{\protect \APACyear {2018}}%
{\protect \APACexlab {{\protect \BCnt {2}}}}}]{%
stan2018stan}
\APACinsertmetastar {%
stan2018stan}%
\begin{APACrefauthors}%
{Stan Development Team}.%
\end{APACrefauthors}%
\unskip\
\newblock
\APACrefYearMonthDay{2018{\protect \BCnt {2}}}{}{}.
\newblock
{\BBOQ}\APACrefatitle {Stan Modeling Language Users Guide and Reference Manual}
  {Stan modeling language users guide and reference manual}.{\BBCQ}
\newblock
\APACjournalVolNumPages{Version 2.18.0}{}{}{}.
\newblock
\APAChowpublished {\url{http://mc-stan.org}}.
\PrintBackRefs{\CurrentBib}

\bibitem [\protect \citeauthoryear {%
Tufto%
\ \protect \BOthers {.}}{%
Tufto%
\ \protect \BOthers {.}}{%
{\protect \APACyear {2012}}%
}]{%
tufto2012estimating}
\APACinsertmetastar {%
tufto2012estimating}%
\begin{APACrefauthors}%
Tufto, J.%
, Lande, R.%
, Ringsby, T\BHBI H.%
, Engen, S.%
, S{\ae}ther, B\BHBI E.%
, Walla, T\BPBI R.%
\BCBL {}\ \BBA {} DeVries, P\BPBI J.%
\end{APACrefauthors}%
\unskip\
\newblock
\APACrefYearMonthDay{2012}{}{}.
\newblock
{\BBOQ}\APACrefatitle {Estimating {B}rownian motion dispersal rate, longevity
  and population density from spatially explicit mark--recapture data on
  tropical butterflies} {Estimating {B}rownian motion dispersal rate, longevity
  and population density from spatially explicit mark--recapture data on
  tropical butterflies}.{\BBCQ}
\newblock
\APACjournalVolNumPages{Journal of Animal Ecology}{81}{4}{756--769}.
\PrintBackRefs{\CurrentBib}

\bibitem [\protect \citeauthoryear {%
Tufto%
\ \protect \BOthers {.}}{%
Tufto%
\ \protect \BOthers {.}}{%
{\protect \APACyear {2000}}%
}]{%
tufto2000bayesian}
\APACinsertmetastar {%
tufto2000bayesian}%
\begin{APACrefauthors}%
Tufto, J.%
, S{\ae}ther, B\BHBI E.%
, Engen, S.%
, Arcese, P.%
, Jerstad, K.%
, R{\o}stad, O\BPBI W.%
\BCBL {}\ \BBA {} Smith, J\BPBI N.%
\end{APACrefauthors}%
\unskip\
\newblock
\APACrefYearMonthDay{2000}{}{}.
\newblock
{\BBOQ}\APACrefatitle {Bayesian meta-analysis of demographic parameters in
  three small, temperate passerines} {Bayesian meta-analysis of demographic
  parameters in three small, temperate passerines}.{\BBCQ}
\newblock
\APACjournalVolNumPages{Oikos}{88}{2}{273--281}.
\PrintBackRefs{\CurrentBib}

\bibitem [\protect \citeauthoryear {%
Vedder%
, Bouwhuis%
\BCBL {}\ \BBA {} Sheldon%
}{%
Vedder%
\ \protect \BOthers {.}}{%
{\protect \APACyear {2013}}%
}]{%
vedder2013quantitative}
\APACinsertmetastar {%
vedder2013quantitative}%
\begin{APACrefauthors}%
Vedder, O.%
, Bouwhuis, S.%
\BCBL {}\ \BBA {} Sheldon, B\BPBI C.%
\end{APACrefauthors}%
\unskip\
\newblock
\APACrefYearMonthDay{2013}{}{}.
\newblock
{\BBOQ}\APACrefatitle {Quantitative assessment of the importance of phenotypic
  plasticity in adaptation to climate change in wild bird populations}
  {Quantitative assessment of the importance of phenotypic plasticity in
  adaptation to climate change in wild bird populations}.{\BBCQ}
\newblock
\APACjournalVolNumPages{PLoS Biology}{11}{7}{e1001605}.
\newblock
\begin{APACrefDOI} \doi{10.1371/journal.pbio.1001605} \end{APACrefDOI}
\PrintBackRefs{\CurrentBib}

\bibitem [\protect \citeauthoryear {%
Wei%
}{%
Wei%
}{%
{\protect \APACyear {2006}}%
}]{%
wei2006time}
\APACinsertmetastar {%
wei2006time}%
\begin{APACrefauthors}%
Wei, W.%
\end{APACrefauthors}%
\unskip\
\newblock
\APACrefYear{2006}.
\newblock
\APACrefbtitle {Time Series Analysis: Univariate and Multivariate Methods}
  {Time series analysis: Univariate and multivariate methods}\
  (\PrintOrdinal{2nd}\ \BEd).
\newblock
\APACaddressPublisher{}{Pearson Addison Wesley}.
\PrintBackRefs{\CurrentBib}

\bibitem [\protect \citeauthoryear {%
Zellner%
}{%
Zellner%
}{%
{\protect \APACyear {1986}}%
}]{%
zellner1986assessing}
\APACinsertmetastar {%
zellner1986assessing}%
\begin{APACrefauthors}%
Zellner, A.%
\end{APACrefauthors}%
\unskip\
\newblock
\APACrefYearMonthDay{1986}{}{}.
\newblock
{\BBOQ}\APACrefatitle {On assessing prior distributions and {B}ayesian
  regression analysis with g-prior distributions} {On assessing prior
  distributions and {B}ayesian regression analysis with g-prior
  distributions}.{\BBCQ}
\newblock
\APACjournalVolNumPages{Bayesian inference and decision techniques}{}{}{}.
\PrintBackRefs{\CurrentBib}

\end{thebibliography}

\end{document}


\section*{Supporting Information (SI) for}
\subsubsection*{\bf Bayesian inference with tmbstan for a state-space model with VAR(1) state equation}

\section{Supplementary results of simulation studies}
\begin{table}[]
\centering
\caption{Frequentsit and Bayesian estimates from the model with AR(1) $\theta_t$, autocorrelation in $\theta_t$ $\phi_{\theta,\theta}=0.1$, and different sample sizes.}
\label{table:S1}
\begin{tabular}{|l|l|l|l|l|} \hline 
\multicolumn{5}{|c|}{$\phi_{\theta,\theta}=0.1$, $tmax=25$, $n=50$} \\ \hline 
Parameters & True value & MLE & Prior1 & Prior2  \\ \hline 
no. divergent transitions & NA & NA &1  &  0 \\ \hline 
$\mu_\alpha$ & 2 &2.006(0.016) & 2.005(0.016) & 2.006(0.016) \\ \hline 
$\mu_\theta$ & 20 &19.6(6.4) & 19.3(9.3) & 19.8(6.1) \\ \hline 
$\mu_\omega$ & 3.5 &3.475(0.030) & 3.479(0.030) & 3.472(0.030) \\ \hline 
$\phi_{\theta,\theta}$ & 0.1 &0.26(0.19) & 0.34(0.22) & 0.25(0.16) \\ \hline 
$log\sigma_\theta$ & 2.996 &3.21(0.16) & 3.34(0.20) & 3.11(0.14) \\ \hline 
\multicolumn{5}{|c|}{$\phi_{\theta,\theta}=0.1$, $tmax=25$, $n=100$} \\ \hline 
Parameters & True value & MLE & Prior1 & Prior2  \\ \hline 
no. divergent transitions & NA & NA &1  &  0 \\ \hline 
$\mu_\alpha$ & 2 &1.996(0.010) & 1.996(0.010) & 1.997(0.010) \\ \hline 
$\mu_\theta$ & 20 &17.1(3.7) & 16.4(5.0) & 17.0(3.8) \\ \hline 
$\mu_\omega$ & 3.5 &3.493(0.021) & 3.494(0.022) & 3.491(0.022) \\ \hline 
$\phi_{\theta,\theta}$ & 0.1 &0.07(0.21) & 0.15(0.24) & 0.10(0.18) \\ \hline 
$log\sigma_\theta$ & 2.996 &2.85(0.15) & 2.95(0.18) & 2.78(0.13) \\ \hline 
\multicolumn{5}{|c|}{$\phi_{\theta,\theta}=0.1$, $tmax=50$, $n=25$} \\ \hline 
Parameters & True value & MLE & Prior1 & Prior2  \\ \hline 
no. divergent transitions & NA & NA &0  &  0 \\ \hline 
$\mu_\alpha$ & 2 &1.977(0.015) & 1.977(0.015) & 1.976(0.015) \\ \hline 
$\mu_\theta$ & 20 &19.8(2.7) & 19.7(3.1) & 19.9(2.8) \\ \hline 
$\mu_\omega$ & 3.5 &3.529(0.033) & 3.535(0.033) & 3.525(0.033) \\ \hline 
$\phi_{\theta,\theta}$ & 0.1 &0.04(0.15) & 0.07(0.17) & 0.06(0.14) \\ \hline 
$log\sigma_\theta$ & 2.996 &2.88(0.12) & 2.93(0.12) & 2.83(0.12) \\ \hline 
\multicolumn{5}{|c|}{$\phi_{\theta,\theta}=0.1$, $tmax=50$, $n=100$} \\ \hline 
Parameters & True value & MLE & Prior1 & Prior2  \\ \hline 
no. divergent transitions & NA & NA &0  &  0 \\ \hline 
$\mu_\alpha$ & 2 &1.9858(0.0076) & 1.9857(0.0078) & 1.9856(0.0077) \\ \hline 
$\mu_\theta$ & 20 &20.3(2.8) & 20.3(2.9) & 20.3(2.9) \\ \hline 
$\mu_\omega$ & 3.5 &3.515(0.015) & 3.515(0.016) & 3.513(0.015) \\ \hline 
$\phi_{\theta,\theta}$ & 0.1 &0.09(0.14) & 0.12(0.16) & 0.11(0.14) \\ \hline 
$log\sigma_\theta$ & 2.996 &2.89(0.10) & 2.93(0.11) & 2.86(0.10) \\ \hline 
\end{tabular}
\end{table}

\begin{table}[]
\centering
\caption{Frequentsit and Bayesian estimates from the model with AR(1) $\theta_t$, autocorrelation in $\theta_t$ $\phi_{\theta,\theta}=0.7$, and different sample sizes.}
\label{table:S2}
\begin{tabular}{|l|l|l|l|l|} \hline 
\multicolumn{5}{|c|}{$\phi_{\theta,\theta}=0.7$, $tmax=25$, $n=50$} \\ \hline 
Parameters & True value & MLE & Prior1 & Prior2  \\ \hline 
no. divergent transitions & NA & NA &1  &  0 \\ \hline 
$\mu_\alpha$ & 2 &2.012(0.015) & 2.012(0.015) & 2.011(0.014) \\ \hline 
$\mu_\theta$ & 20 &15.7(4.8) & 15.0(8.7) & 16.0(5.0) \\ \hline 
$\mu_\omega$ & 3.5 &3.483(0.031) & 3.486(0.031) & 3.480(0.031) \\ \hline 
$\phi_{\theta,\theta}$ & 0.7 &0.45(0.18) & 0.55(0.19) & 0.43(0.16) \\ \hline 
$log\sigma_\theta$ & 2.996 &2.72(0.18) & 2.89(0.27) & 2.65(0.16) \\ \hline 
\multicolumn{5}{|c|}{$\phi_{\theta,\theta}=0.7$, $tmax=25$, $n=100$} \\ \hline 
Parameters & True value & MLE & Prior1 & Prior2  \\ \hline 
no. divergent transitions & NA & NA &NA  &  0 \\ \hline 
$\mu_\alpha$ & 2 &1.987(0.011) & 1.980(0.014) & 1.986(0.011) \\ \hline 
$\mu_\theta$ & 20 &18.3(9.7) & 20(18) & 18.4(8.1) \\ \hline 
$\mu_\omega$ & 3.5 &3.539(0.022) & 3.566(0.049) & 3.537(0.022) \\ \hline 
$\phi_{\theta,\theta}$ & 0.7 &0.70(0.13) & 0.60(0.35) & 0.63(0.11) \\ \hline 
$log\sigma_\theta$ & 2.996 &3.10(0.23) & 3.36(0.29) & 2.95(0.16) \\ \hline 
\multicolumn{5}{|c|}{$\phi_{\theta,\theta}=0.7$, $tmax=50$, $n=25$} \\ \hline 
Parameters & True value & MLE & Prior1 & Prior2  \\ \hline 
no. divergent transitions & NA & NA &1  &  0 \\ \hline 
$\mu_\alpha$ & 2 &2.021(0.016) & 2.021(0.016) & 2.021(0.016) \\ \hline 
$\mu_\theta$ & 20 &19.3(8.9) & 20(14) & 19.7(7.8) \\ \hline 
$\mu_\omega$ & 3.5 &3.488(0.031) & 3.490(0.030) & 3.482(0.030) \\ \hline 
$\phi_{\theta,\theta}$ & 0.7 &0.739(0.094) & 0.781(0.091) & 0.692(0.081) \\ \hline 
$log\sigma_\theta$ & 2.996 &3.24(0.18) & 3.39(0.25) & 3.13(0.14) \\ \hline 
\multicolumn{5}{|c|}{$\phi_{\theta,\theta}=0.7$, $tmax=50$, $n=100$} \\ \hline 
Parameters & True value & MLE & Prior1 & Prior2  \\ \hline 
no. divergent transitions & NA & NA &1  &  0 \\ \hline 
$\mu_\alpha$ & 2 &1.9899(0.0076) & 1.9899(0.0076) & 1.9896(0.0075) \\ \hline 
$\mu_\theta$ & 20 &21.1(6.2) & 20(12) & 21.6(5.5) \\ \hline 
$\mu_\omega$ & 3.5 &3.511(0.015) & 3.511(0.015) & 3.510(0.015) \\ \hline 
$\phi_{\theta,\theta}$ & 0.7 &0.71(0.10) & 0.76(0.10) & 0.667(0.086) \\ \hline 
$log\sigma_\theta$ & 2.996 &2.93(0.17) & 3.09(0.27) & 2.84(0.14) \\ \hline 
\end{tabular}
\end{table}

Similar to Table 1 in the main text, we here show the frequentist and Bayesian estimates of the same parameters but with different true
values of $\phi_{\theta,\theta}$. Table \ref{table:S1} and Table \ref{table:S2} list the estimates of parameters under different simulation
settings with $\phi_{\theta,\theta}=0.1$ and 0.7 respectively. From these two tables, we find generally similar patterns 
to the table of estimates in the main text. For example, dataset with more time points $(tmax=50, n=100)$  leads to more accurate estimates compared with the dataset with shorter time series $(tmax=25, n=100)$. Increasing the sample size at each time point improves neither the accuracy nor the certainty of the estimates for the parameters of interest, only a bigger sample size is required for this purpose.

\begin{figure}%
    \centering
    \subfloat[$(tmax=25,n=50)$ with Prior1.]{{\includegraphics[width=9cm]{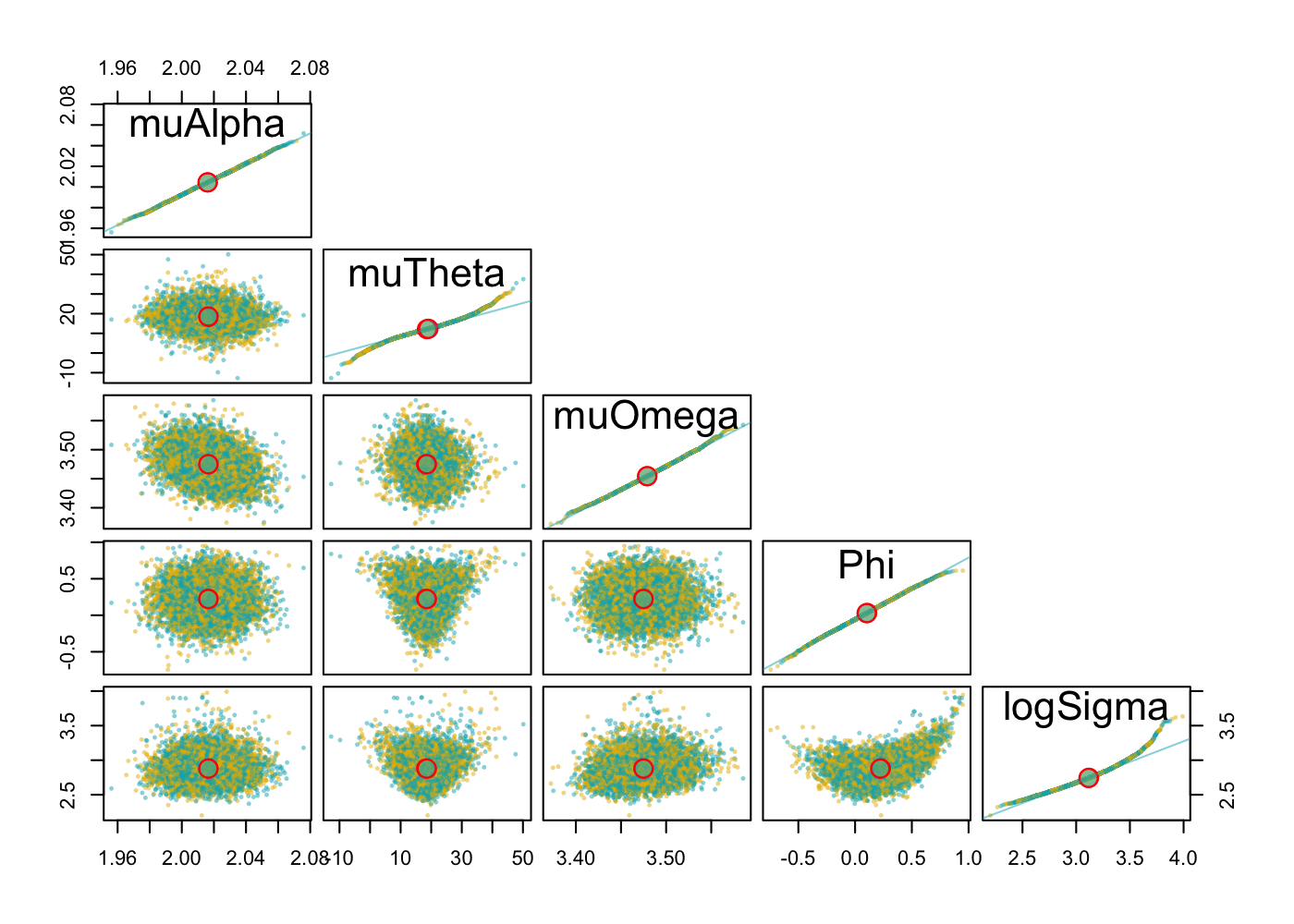}}}%
    \subfloat[$(tmax=25,n=100)$ with Prior1.]{{\includegraphics[width=9cm]{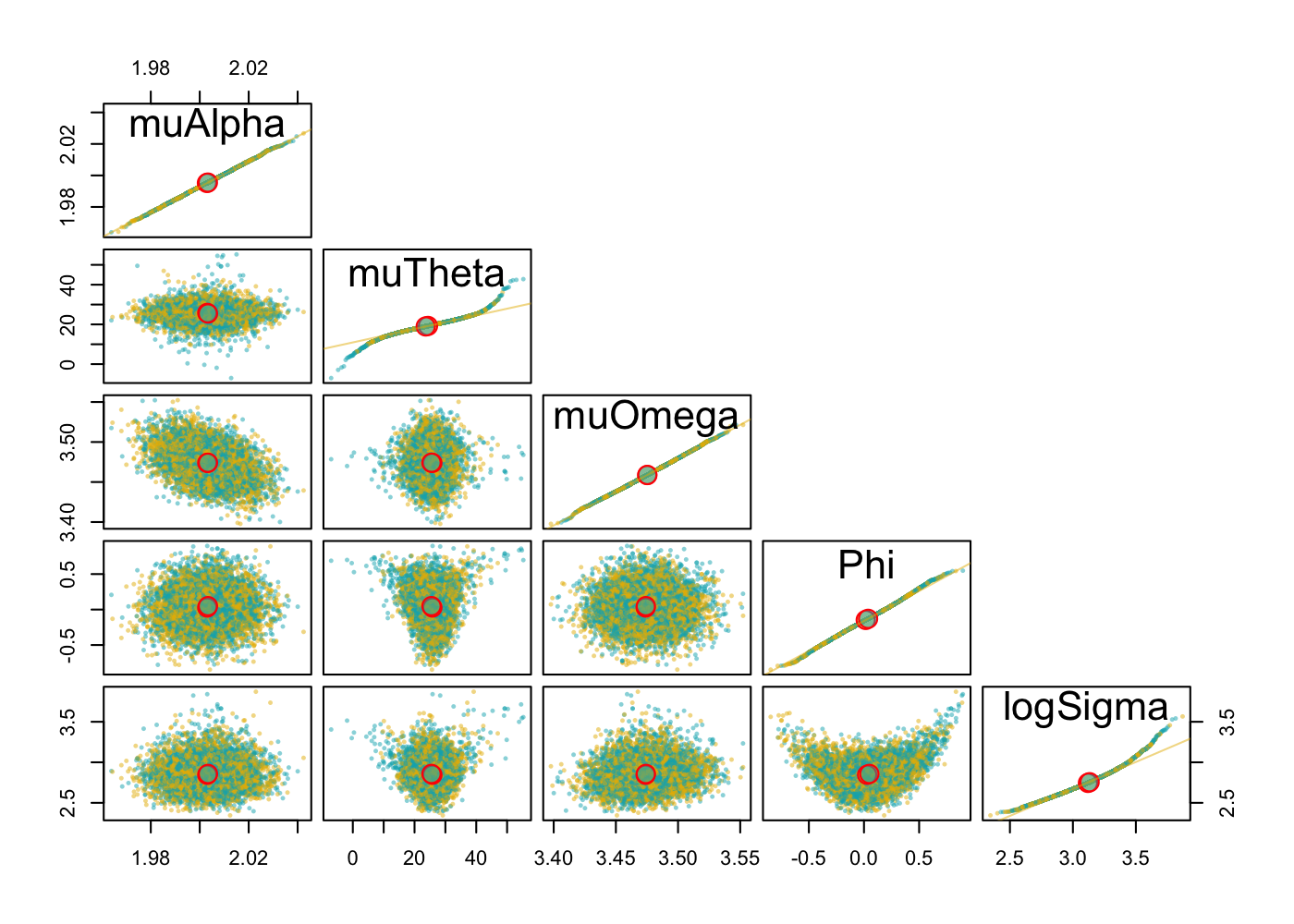}}}%
        \qquad
    \subfloat[$(tmax=50,n=25)$ with Prior1.]{{\includegraphics[width=9cm]{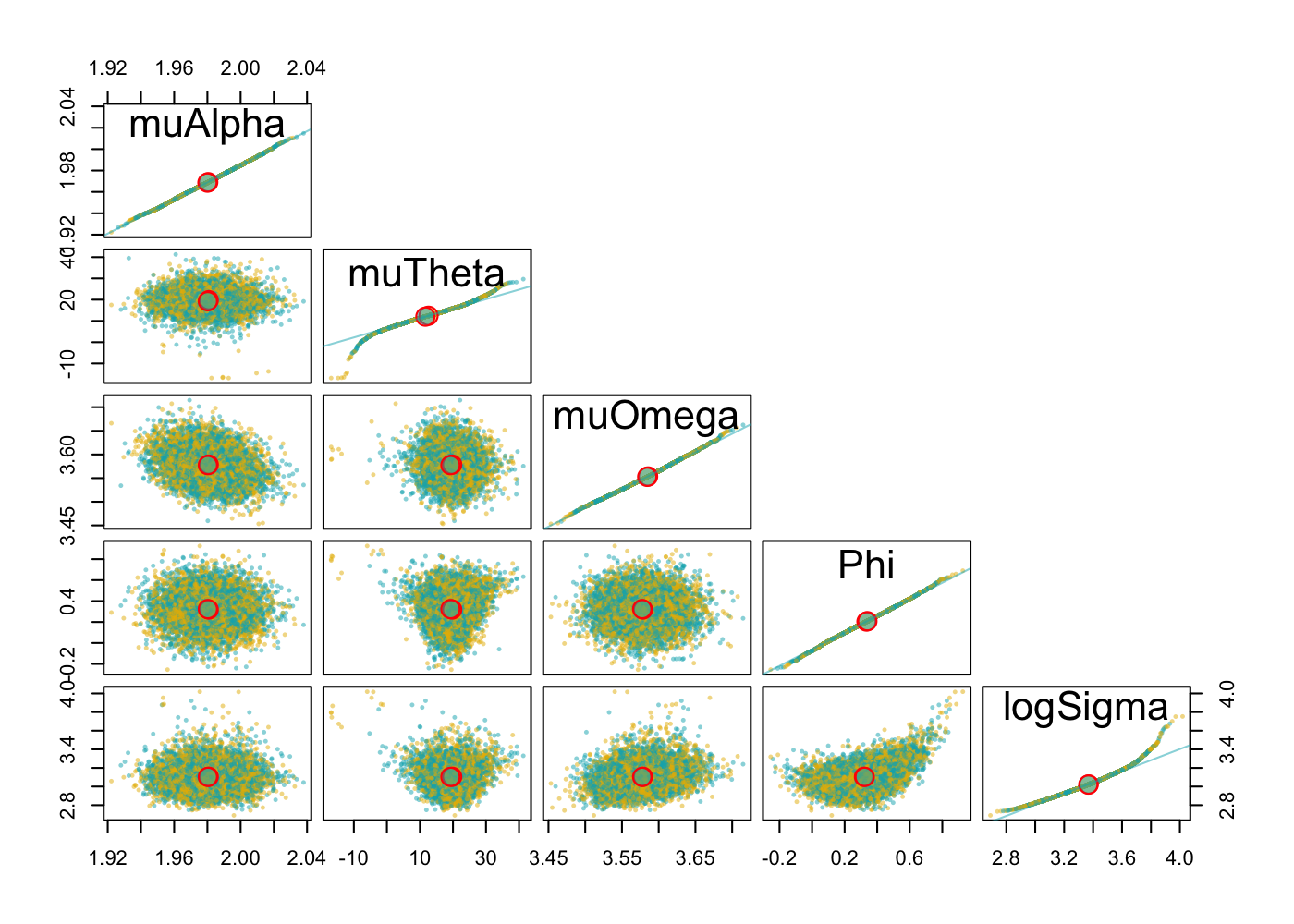}}}%
    \subfloat[$(tmax=50,n=100)$ with Prior1.]{{\includegraphics[width=9cm]{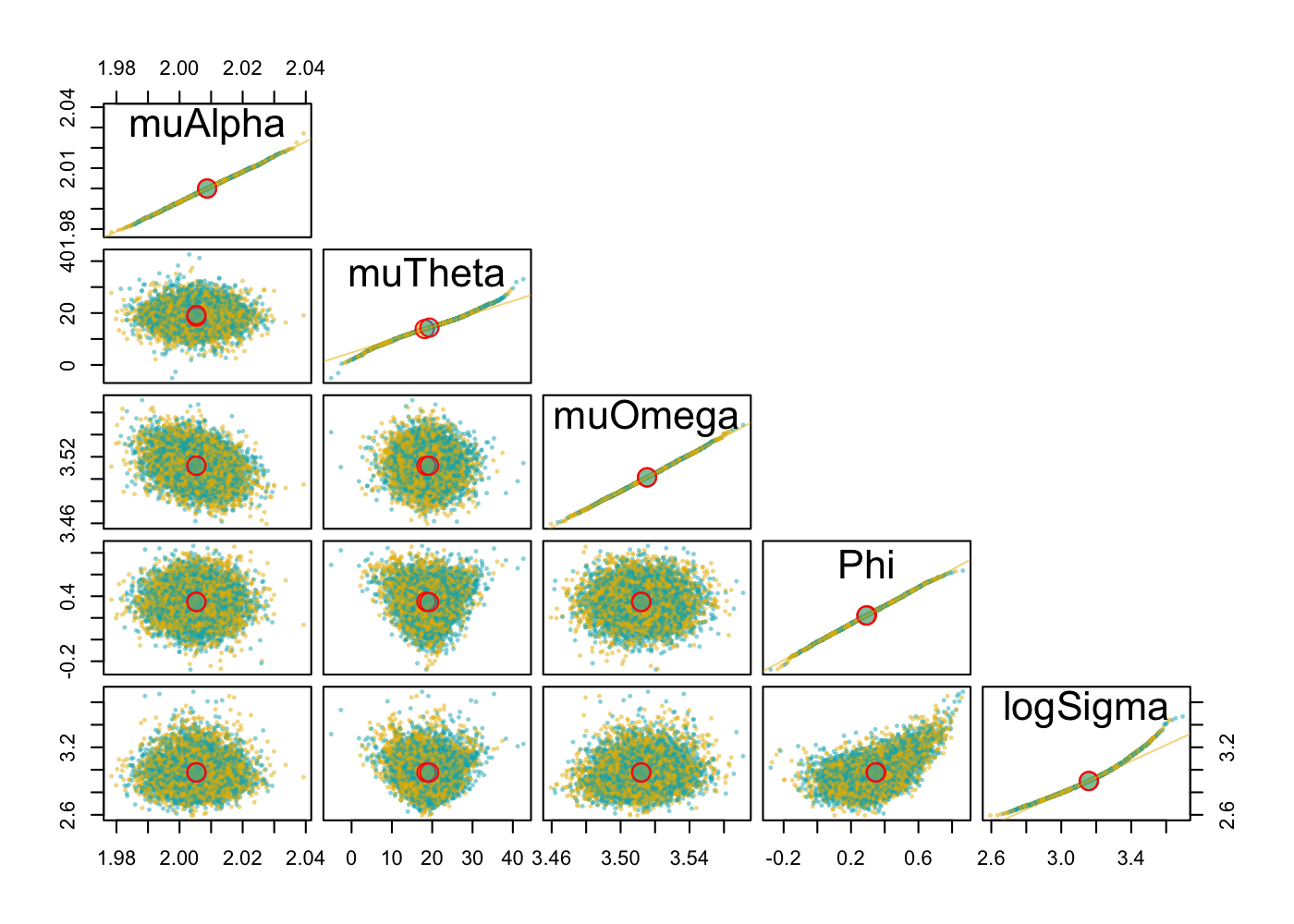}}}%
    \caption{Pair plots of posterior samples for Laplace approximation check for one realization of the simulation with prior1. 
    The four plots (a), (b), (c), and (d) correspond to the four schemes of simulation. The random effects in the TMB model can be integrated with two techniques: (1) full MCMC integration via NUTS and  (2) Laplace approximation. To check the accuracy of Laplace approximation to the posterior likelihood density, the posterior samples for all the fixed effects in the model without (yellow dots) and with Laplace approximation (green dots) are shown pair-wisely on the same plot. Columns and rows on the lower diagonal correspond to pair-wise parameters, with the diagonal showing QQ-plot of posterior samples from Bayesian inference without (yellow dots) and with (green dots) for that parameter including a 1:1 line in yellow. The large red circles of the off-diagonal plots represent the pairwise means. On each off-diagonal plot, there are 4000 yellow dots corresponding to 1000 samples retained from each of four chains without Laplace approximation, so as to the green dots with Laplace approximation. Posterior rows were randomized to prevent consistent overplotting of one integration technique. Overlaps in the two colored dots suggest the Laplace approximation assumption is met.}
    \label{fig:lacheckP1}%
\end{figure}

\begin{figure}%
    \centering
    \rowname{Prior1}
    \subfloat[$(tmax=25,n=50)$]{{\includegraphics[width=4cm]{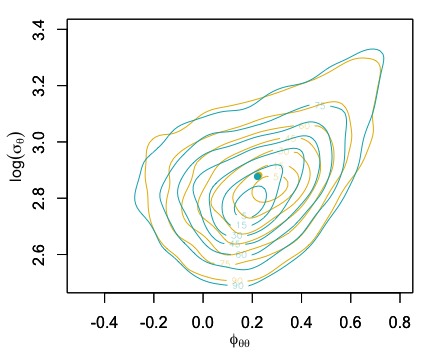}}}
    \subfloat[$(tmax=25,n=100)$]{{\includegraphics[width=4cm]{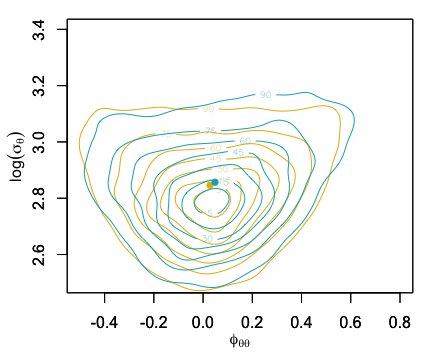}}}
    \subfloat[$(tmax=50,n=25)$]{{\includegraphics[width=4cm]{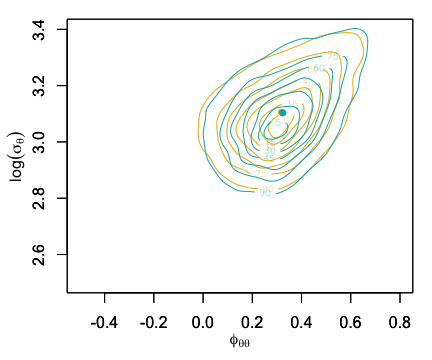}}}
    \subfloat[$(tmax=50,n=100)$]{{\includegraphics[width=4cm]{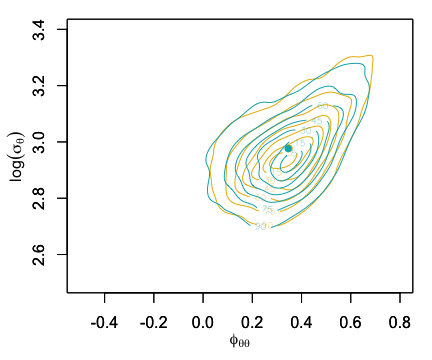}}} \\[2.5ex] 
    \rowname{Prior2}
    \subfloat[$(tmax=25,n=50)$]{{\includegraphics[width=4cm]{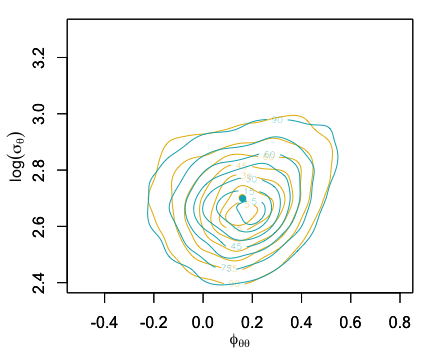}}}
    \subfloat[$(tmax=25,n=100)$]{{\includegraphics[width=4cm]{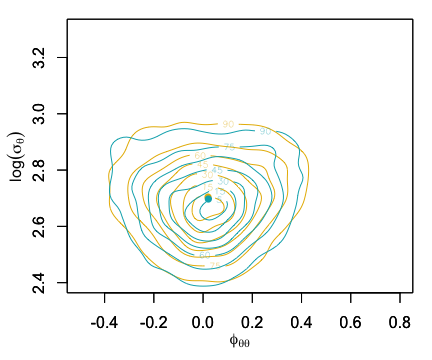}}}
    \subfloat[$(tmax=50,n=25)$]{{\includegraphics[width=4cm]{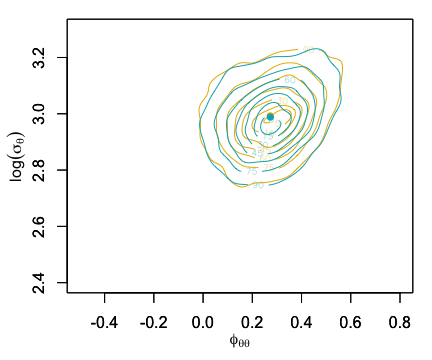}}}
    \subfloat[$(tmax=50,n=100)$]{{\includegraphics[width=4cm]{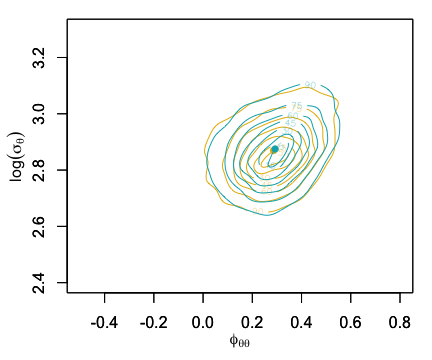}}}
    \caption{Bivariate contour plots of posterior samples of $\phi_{\theta,\theta}$ and $\log(\sigma_{\theta})$ from one realization of the simulation with Prior1 (the first row) and Prior2 (the second row) for Laplace approximation check. The posterior samples data used are the same as that in Fig.~\ref{fig:lacheckP1} and Figure 4 in the main text. The yellow contours indicate the joint posterior distribution of ($\phi_{\theta,\theta}$, $\log(\sigma_{\theta})$) from the estimation technique full MCMC integration via NUTS, and the green contours correspond to the technique that Laplace approximation is used. The yellow and green dots in each plot represent the mean of the bivariate posterior samples in each setting respectively.}
    \label{fig:lacheckContour}%
\end{figure}

In the main text, we only present the pair plots of posterior samples for Laplace approximation check with Prior2. We here supplement the pair plots (Fig.~\ref{fig:lacheckP1}) with Prior1 under the four different sample size settings. Fig.~\ref{fig:lacheckP1} also suggests accurate Laplace approximation indicated by the good mix of posterior samples. To further validate this conclusion, we visually inspect the accuracy of Laplace approximation by plotting bivariate contour plots of posterior samples from the Bayesian model with and without Laplace approximation on the same figure, as shown in  Fig.~\ref{fig:lacheckContour}. Only the joint posterior distribution ($\phi_{\theta,\theta}$ and $\log(\sigma_{\theta})$) is considered and other parameters are ignored for simplifying the analysis. The overlap of contours with (yellow) and without (green) Laplace approximation for the random effects suggests again that the Laplace approximation in these cases is accurate.

\newpage
\section{Supplementary info on real data case study}
\begin{figure}%
    \centering
    \subfloat[Prior density function on $\sigma_\alpha$ or $\sigma_\theta$ given Inverse-gamma (1, 1) prior on $\sigma_\alpha^2$ or $\sigma_\theta^2$ respectively.]{{\includegraphics[width=10cm]{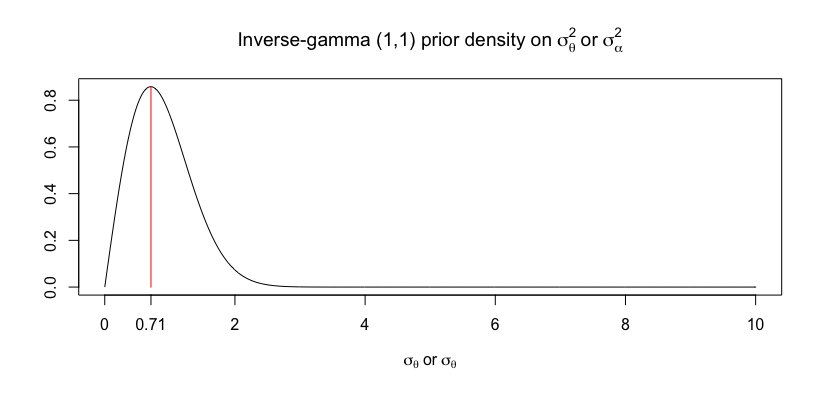} }}%
    \qquad
    \subfloat[Trace plots for $log\sigma_{\alpha}$ (left) and for $log\sigma_{\theta}$ (right).] {{\includegraphics[width=13cm]{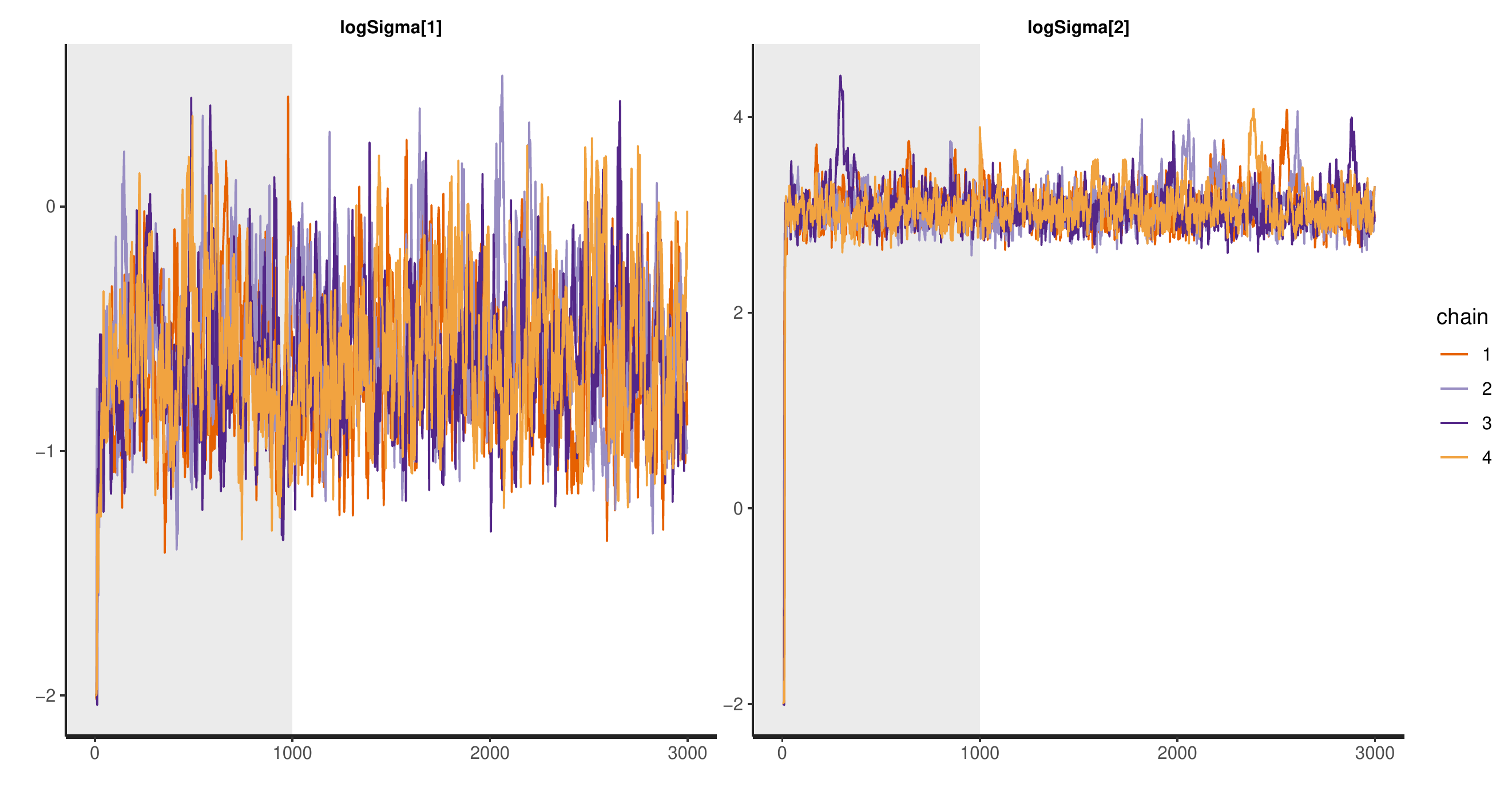} }}%
    \caption{A prior density and trace plots for the great tit case study. In plot (a), the solid curve indicates an equivalence of the density to inverse-gamma (1, 1) prior on $\sigma_{\alpha}^2$ or $\sigma_{\theta}^2$, the equivalent density on $\sigma_{\alpha}$ or $\sigma_{\theta}$ is calculated with rules of density function transformation, which is omitted here. The red solid line indicates the density mode. Plot (b) shows trace plots with the inverse-gamma (1,1) priors for parameter $\sigma_{\alpha}^2$ (left) and $\sigma_{\theta}^2$ (right) respectively. The grey areas indicate warm-up iterations.}%
    \label{fig:traceplot}%
\end{figure}

Beside half-Cauchy and lognormal priors for the scale parameters of the great tit model as shown in the main text, we also tested inverse-gamma (1, 1) prior for the scale parameter $\sigma_\alpha^2$ and  $\sigma_\theta^2$. To visualize MCMC diagnostics we show trace plots for the two scale parameters along with the prior densities in Fig.~\ref{fig:traceplot}. The solid black line in plot (a) shows prior density function of $\sigma_\alpha$
(or $\sigma_\theta$) given a Inverse-gamma (1, 1) prior density on $\sigma_{\alpha}^2$ (or $\sigma_{\theta}^2$). The details on density function transformation are omitted here. The solid red line indicates the density mode. The prior density mode 
of $\sigma_{\alpha}$ at 0.71 translates to density mode of log$\sigma_{\alpha}$ at -0.34. However, the left trace plot in plot (b) for log$\sigma_{\alpha}$ implies that the posterior likelihood is dominated by the prior so that the sampler gets trapped in the subspace of the parameter, which is a space near -0.34, while the true posterior density mode locates around -1.7.

As mentioned in the main text, the great tit model implemented with Bayesian inference was selected in the frequentist framework with model selection
procedure. Table \ref{table:modSelection} lists all the candidate models fitted with the great tit data. Model 7 is selected as the best model due to the smallest AIC value reported. Colomun $\Delta$p and $\Delta$AIC lists the difference between the selected model and
 the corresponding candidate model in the number of parameters and reported AIC value respectively. The rightmost column describes the candidate models.
 \begin{table}[]
 \centering
 \caption{Model selection for the real data case study. The table lists all the candidate models fitted with the great tit data. Model 7 is selected as the best model due to the smallest AIC value. Column $\Delta$p and $\Delta$AIC lists the difference between the selected model and
 the corresponding candidate model in the number of parameters and reported AIC value respectively. The rightmost column describes the candidate models. The elements in matrix $\mathbf{\Phi}$ and vector $\rho$ are set to 0 if not otherwise specified.}
 \label{table:modSelection}
\begin{tabular}{|l|l|l|l|}
\hline
Model & $\Delta$p & $\Delta$AIC & Description \\ \hline
1 &   -5 & 821.51 &  $\alpha_t=\theta_t=\omega_t=0$ \\ \hline
2 &   -4 & 295.07 &  $\theta_t=\omega_t=0$, random $\alpha_t$ \\ \hline
3 &   -3 & 34.81 & $\omega_t=0$, random $\alpha_t$ and $\theta_t$  \\ \hline
4 &   -2 & 176.51 & random $\alpha_t$, $\theta_t$ and $\omega_t$  \\ \hline
5 &   -4 & 265.32 &  $\alpha_t=\omega_t=0$, random $\theta_t$ \\ \hline
6 &  2 & 2.52 & $\omega_t=0$, VAR(1) $\alpha_t$ and $\theta_t$: $\phi_{\alpha,\alpha}$ $\neq$  $\phi_{\theta,\theta}$ $\neq$
$\phi_{\alpha,\theta}$  $\neq$  $\phi_{\theta,\alpha}$ $\neq$ 0, $\rho_{\alpha,\theta}$ $\neq$ 0 \\ \hline
7 (best model) &  0 & 0  & $\omega_t=0$, AR(1) $\alpha_t$ and AR(1) $\theta_t$: $\phi_{\alpha,\alpha}$ $\neq$  $\phi_{\theta,\theta}$ $\neq$ 0  \\ \hline
8 &  1 & 1.92 & $\omega_t=0$, VAR(1) $\alpha_t$ and $\theta_t$: $\phi_{\alpha,\alpha}$ $\neq$  $\phi_{\theta,\theta}$  $\neq$  $\phi_{\theta,\alpha}$ $\neq$ 0,  $\rho_{\alpha,\theta}$ $\neq$ 0   \\ \hline
9 &  1& 1.21 & $\omega_t=0$, VAR(1) $\alpha_t$ and $\theta_t$:  $\phi_{\alpha,\alpha}$ $\neq$  $\phi_{\theta,\theta}$ $\neq$  $\phi_{\alpha,\theta}$   $\neq$ 0, $\rho_{\alpha,\theta}$ $\neq$ 0 \\\hline
10 &  -1 & 12.93 &  $\omega_t=0$, random $\theta_t$, AR(1) $\alpha_t$: $\phi_{\alpha,\alpha}$ $\neq$ 0\\ \hline
11 &  -1 & 6.7 &  $\omega_t=0$, random $\alpha_t$, AR(1) $\theta_t$: $\phi_{\theta,\theta}$ $\neq$ 0 \\ \hline
\end{tabular}
\end{table}

\begin{figure}%
    \centering
    \rowname{(a) Bivariate contour plots with Prior1.}
    \subfloat{{\includegraphics[width=5cm]{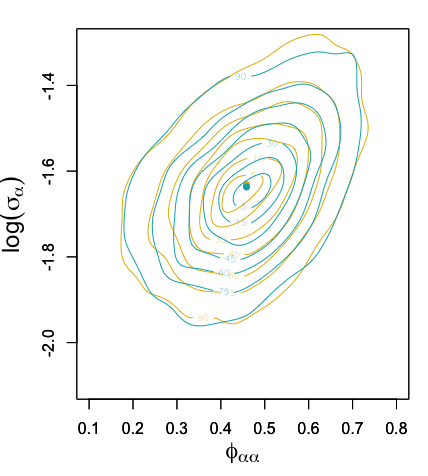}}}
    \subfloat{{\includegraphics[width=5cm]{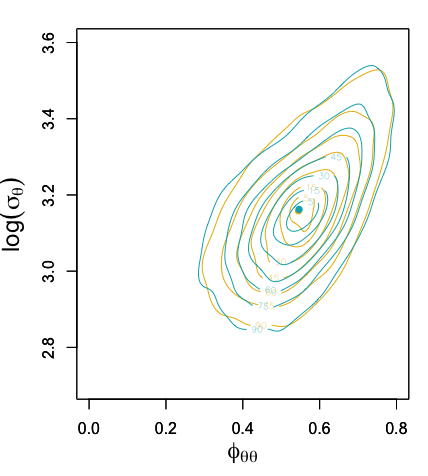}}}
    \subfloat{{\includegraphics[width=5cm]{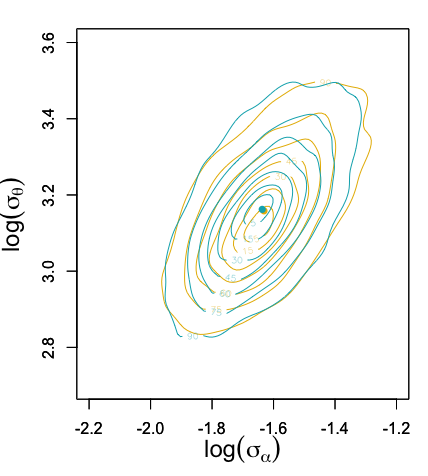}}}
     \\[2.5ex] 
    \quad
    \rowname{(b) Bivariate contour plots with Prior2.}
    \subfloat{{\includegraphics[width=5cm]{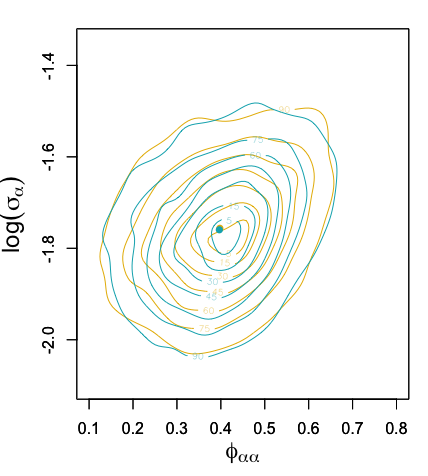}}}
    \subfloat{{\includegraphics[width=5cm]{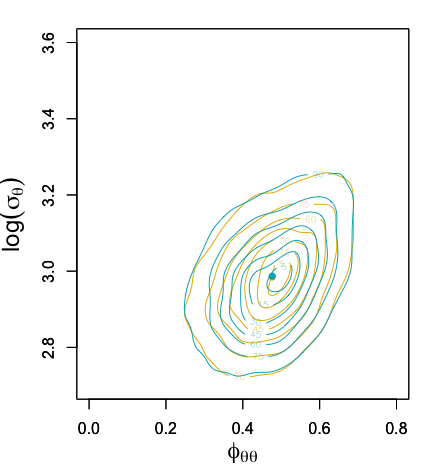}}}
    \subfloat{{\includegraphics[width=5cm]{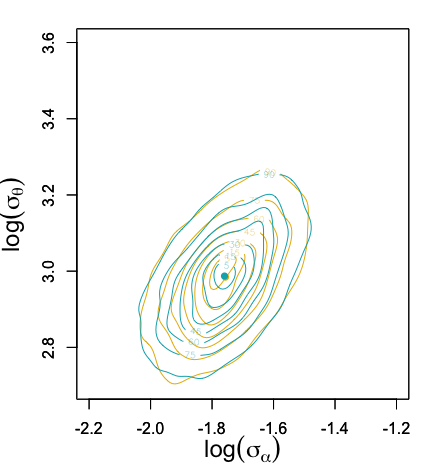}}}
    \caption{Bivariate contour plots of posterior samples of a subset of the parameters in the selected great tit model for Laplace approximation check. The posterior samples used here are the same as that in Figure 5 and Figure 6 in the main text. The plots in row (a) correspond to the Bayesian model with Prior1, and in row (b) they are with Prior2. Similar to Fig.~\ref{fig:lacheckContour}, the yellow contours  indicate the joint posterior distribution of the parameters from the estimation technique full MCMC integration via NUTS, and the green contours correspond to the technique that Laplace approximation is used. The yellow and green dots in each plot again represent the mean of the bivariate posterior samples in each plot respectively. Only a subset of the parameters is considered for simplification. }
    \label{fig:lacheckContourGT}%
\end{figure}

We also plot the contours of posterior samples with and without Laplace approximation for a subset of parameters in the great tit model on the same graph (Fig.~\ref{fig:lacheckContourGT}), to get a clearer visualization of the posteriors' distribution. The first and second row of the contour plots corresponds to the Bayesian great tit model with Prior1 and Prior2 respectively. The round dots on the plots are the mean of posterior samples for each estimation technique. The good amount of overlap of the yellow contours, dots (without Laplace approximation), and green contours, dots (with Laplace approximation) again suggests a good accuracy of Laplace approximation.